\documentclass[%
 amsmath,amssymb, aps, pre, twocolumn
]{revtex4}

\usepackage{amsmath}
\usepackage{graphicx,subfigure,amssymb,verbatim}
\newcommand{\Lapq}{\mathcal{L}_q}
\newcommand{\pofl}{P(\ell)}
\newcommand\distr{\mathrel{\overset{\makebox[0pt]{\mbox{\normalfont\small\sffamily d}}}{=}}}

\newcommand{\RN}[1]{%
  \textup{\uppercase\expandafter{\romannumeral#1}}%
}

\usepackage{color}

\usepackage[usenames,dvipsnames]{xcolor}

\usepackage{color}
\definecolor{darkgreen}{rgb}{0,0.6,0}
\definecolor{darkblue}{rgb}{0,0,0.6}
\definecolor{darkred}{rgb}{0.6,0,0}
\definecolor{darkpurple}{rgb}{0.5,0,0.5}

\usepackage{hyperref}

\hypersetup{
bookmarksopen=true,
pdftitle=Power countings versus physical scalings in 1D DES,
pdfauthor=E Agoritsas and V Lecomte, 
pdftoolbar=false, 
pdfstartview={FitH},		
pdfmenubar=true,			
pdfhighlight=/O,			
colorlinks=true,			
urlcolor=darkblue,
citecolor=darkblue,		
linkcolor=darkpurple	
}

\usepackage{soul}

\begin{document}

\title{Statistics of zero crossings 
in rough interfaces with fractional elasticity}%

\author{Arturo L. Zamorategui}
 \affiliation{Laboratoire Probabilit\'es et Mod\`eles Al\'eatoires (UMR CNRS 7599), Universit\'e Pierre et Marie Curie and Universit\'e Paris Diderot, 75013 Paris, France}
\author{Vivien Lecomte}
\affiliation{Universit\'e Grenoble Alpes, CNRS, LIPhy, 38000 Grenoble, France}
\author{Alejandro B. Kolton}
\affiliation{CONICET and Instituto Balseiro (UNCu), Centro At\'omico Bariloche, 8400 S.C. de Bariloche, Argentina}
\date{\today}
             
\begin{abstract}
We study numerically 
the distribution of zero crossings 
in one-dimensional elastic interfaces described 
by an overdamped Langevin dynamics with periodic boundary conditions. 
We model the 
elastic forces 
with a Riesz-Feller fractional Laplacian 
of order $z = 1 + 2\zeta$, such that 
the interfaces spontaneously relax, with a dynamical exponent $z$, 
to a self-affine geometry with roughness exponent $\zeta$.
By continuously increasing from $\zeta=-1/2$ 
(macroscopically flat interface described by independent Ornstein--Uhlenbeck processes~\cite{uhlembeck1930brownian})
to $\zeta=3/2$ (super-rough Mullins-Herring interface), three 
different regimes are identified:  (I) $-1/2<\zeta<0$, 
(II) $0<\zeta<1$, 
and  (III) $1<\zeta<3/2$. 
Starting from a flat initial condition, the mean number of zeros 
of the discretized interface (I) decays
exponentially in time and reaches an extensive value in the system size, or
decays as a power-law towards (II) a sub-extensive or (III)
an intensive value. 
In the steady-state, 
the distribution of intervals between zeros changes from an exponential decay in (I) to a
power-law decay $P(\ell) \sim \ell^{-\gamma}$ in (II) and (III).
While in (II) $\gamma=1-\theta$ with $\theta=1-\zeta$ the steady-state persistence exponent, in (III) we obtain $\gamma = 3-2\zeta$, different from the exponent $\gamma=1$ expected from the prediction $\theta=0$ for infinite super-rough interfaces with $\zeta>1$.
The effect on $\pofl$ of short-scale smoothening is also analyzed numerically and analytically.
A tight relation between the mean interval, the mean width of the interface and the density of zeros is also reported.
The results drawn from our analysis of rough interfaces subject to particular boundary conditions or constraints, along with discretization effects, are relevant for the practical analysis of zeros in interface imaging experiments or in numerical analysis. 
\end{abstract}

\maketitle

\section{\label{sec:intro}Introduction}

Persistence and first-passage properties in stochastic systems are widely studied in physics, chemistry, biology, finance, engineering~\cite{redner_guide_2001}. The question  
is to predict for how long or how far a certain property remains unchanged, and to determine the probability that such property changes at a certain time $t$ or at a certain position $x$.

In particular, stochastic interfaces
are an interesting case displaying both 
non-trivial temporal and spatial persistence 
properties~\cite{bray_persistence_2013}.
They model, at a coarse grained level, a large variety of 
extended non-equilibrium systems, 
from surface growth by molecular beam epitaxy~\cite{Barabasi-Stanley}, driven domain 
walls in ferromagnetic~\cite{Ferre2013,Ferrero2013} or ferroelectrical materials~\cite{kleemann_review_ferroelectrics}, cracks~\cite{alava_review_cracks,ponson_fracture}, 
growing droplets of turbulent phase 
in nematic liquid crystals~\cite{Takeuchi_2010,Takeuchi2011},
to biological non-equilibrium processes such 
as bacterial colonies~\cite{MATSUSHITA1998517} or tumor growth~\cite{Bru_PhysRevLett.81.4008,BRU20032948}. 
Both the stationary and the aging dynamics of these 
interfaces are experimentally 
relevant. Understanding the universal persistence 
properties in mathematically tractable models of rough interfaces allows a general and quantitative statistical characterization which goes beyond the standard dynamic scaling analysis.

In this paper, we study the transient and steady-state 
\textit{spatial} persistence properties of 
fluctuating one dimensional elastic interfaces 
 described by a univalued 
scalar displacement field $u_x(t)$ at position $x$ and time $t$.
We focus on Gaussian interfaces controlled by the 
linear Langevin equation
\begin{equation}
\partial_t u_x(t)=-(-\partial_x^2)^{z/2}u_x(t)+\eta(x,t), 
\label{eq:modelrealspace}
\end{equation}
where $\eta(x,t)$ is the standard 
non-conserving Gaussian noise with zero mean  and variance 
$\langle \eta(x,t)\eta(x',t') \rangle=2T \delta(x-x')\delta(t-t')$, with $T$ the temperature.
A generalized harmonic elasticity is conveniently 
implemented by using a Riesz-Feller fractional Laplacian 
of order $z$~\cite{zoia_2007}. The order of the Laplacian is the dynamical exponent $z=d+\zeta$ of the interface, where $d$ is the spatial dimension and $\zeta$ the roughness exponent associated.
At zero temperature, Eq.~\eqref{eq:modelrealspace} 
reduces to the well known fractional diffusion equation, $\partial_t u_x(t)=-(-\partial_x^2)^{z/2}u_x(t)$~\cite{METZLER20001}.
In presence of noise and for general $\zeta$, 
Eq.~\eqref{eq:modelrealspace} is a convenient linear but spatially non-local stochastic model that can describe the critical relaxation of an interface towards a self-affine geometry. This simple model allows us to study in an approximate way the geometry of (not necessarily linear nor Gaussian) interfaces in different experimentally relevant universality classes.
For one-dimensional interfaces and $z=2$, Eq.~\eqref{eq:modelrealspace} reduces to 
the well known Edwards-Wilkinson (EW) equation 
describing an elastic string with short-range 
elasticity~\cite{edwards_surface_1982}.   
The fractional Laplacian with $z=1$, models the long-range elasticity 
of contact lines in a liquid meniscus~\cite{ledoussal_contact_line}, 
cracks fronts~\cite{ponson_depinning} or the elastic interactions of magnetic domain walls involved in the 
Barkhausen effect~\cite{Durin_2006}. The case $z=4$ corresponds effectively to restoring elastic forces that depend on the interface curvature~\cite{mullinsherrins}, useful to model surface growth by molecular beam epitaxy in the presence of 
surface diffusion~\cite{Wolf_1990,DasSarma_1991,Barabasi-Stanley}. 
The extreme case $z=0$ represents independent Ornstein--Uhlenbeck 
processes at each position $x$. Our study comprises the continuous range 
of values from $z=0$ to $z=4$, covering macroscopically flat to super-rough interfaces.
We consider 
both the steady-state 
and the transient regimes of 
Eq.~\eqref{eq:modelrealspace}, starting from 
a flat initial condition.

A central quantity to characterize the spatial persistence
of interfaces generated by 
Eq.~\eqref{eq:modelrealspace} is the probability $Q(x_0 , x_0 + x)$ that 
the displacement field $u_x$ 
does not return to its value $u_{x_0}$ 
over the spatial interval $[x_0 , x + x_0 ]$ along
a given direction, at a \textit{fixed} time $t$. 
This property, that has its full 
analog in the temporal persistence of the interface 
at a fixed point $x$, 
is particularly interesting in the long-time limit 
where the stochastic interface develops long-range 
correlations. It is then expected that 
$Q(x_0 , x_0 + x) \sim |x|^{-\theta}$ at large enough $x$, 
with a non-trivial, and probably universal, persistence 
exponent $\theta$. Two independent spatial persistence 
exponents can be found: $\theta=\theta_{\text{FIC}}$ if $x_0$ is initially sampled from the subset 
of points where $\partial^m_x u_x$ is finite for all $m\in\mathbb{N}$, and $\theta=\theta_{\text{SS}}$ if $x_0$ 
is sampled uniformly as studied in~\cite{bray_persistence_2013}.
Majumdar and Bray~\cite{majumdar_spatial_2001} showed 
that $\theta_{\text{SS}}=1-\zeta$ is exact for 
$0<\zeta<1$. Moreover, they showed that the stationary spatial 
persistence properties of Eq.~\eqref{eq:modelrealspace} 
can be mapped to the temporal persistence properties of the 
generalized random-walk process 
$d^nx/dt^n=\eta(t)$ by choosing $n=(z-d+1)/2$, with 
$d$ the dimension of the elastic manifold.
Exploiting this mapping, the authors found that $\theta_{\text{FIC}}=\theta(n)$, 
with $\theta(n)$ the temporal persistence exponent of the 
random-walk process. Thus, the normal Brownian motion for 
$n=1$ corresponds to the one-dimensional Edwards-Wilkinson equation, 
while the so-called random-acceleration process for $n=2$ corresponds 
to the Mullins-Herring equation~\cite{mullinsherrins}.
These two cases are special because they are Markovian, 
and their exponents, $\theta(1)=1/2$ and $\theta(2)=1/4$,
can be exactly computed. For other values of $n$ the equation 
of motion has memory and to estimate $\theta(n)$ we need to rely 
on approximated methods, such as the independent 
interval approximation (IIA)~\cite{derrida_persistent_1996}.

Although exact results are known for infinite continuous interfaces, as described above, it is far from trivial to apply them to finite discrete interfaces. While the former is relevant for theoretical purposes, the latter can be useful for applications to interface imaging experiments. 
In such finite interfaces, the boundary conditions play a fundamental role as it will be soon evident. In systems with long-range elastic interactions, the boundary conditions have to be properly defined, particularly regarding the fractional Laplacian operator in Eq.~\eqref{eq:modelrealspace}~\cite{zoia_2007}. Even in the simplest case of periodic boundary 
conditions (where the steady-states are translational invariant and the Laplacian operator reduces to $-|q|^{z}$ in Fourier space), finite-size effects 
are important~\cite{constantin_spatial_2004, constantin_persistence_2004}.
One of the main issues in the computation of persistence properties in finite-size 
periodic fluctuating interfaces is that the intervals between successive zero crossings are not statistically independent.
For infinite interfaces or temporal signals it is known that the zero crossings of a fractional Brownian motion cannot be accurately described by a renewal process, except for $\zeta=1/2$ which maps 
to normal Markovian Brownian motion ~\cite{reinaldo2010}.
For the $\zeta=1/2$ Edwards-Wilkinson finite interface however, the intervals generated by the crossing zeros of the interface with its center of mass are no longer independent.~\cite{zamorategui16distribution}
Analytical calculations for the survival probability of the EW interface have indeed
shown the importance of the zero-area constraint in finite interfaces~\cite{majumdar_spatial_2006}.
More recently, we have 
shown that the same model displays subtle correlations effects between intervals and long-range correlations between increments~\cite{zamorategui16distribution}.
Thus, finite-size effects are expected to become even more important for fractional dynamics, specially  for large values of $\zeta$ where excursions get even more constrained by the zero-area condition.   
Moreover, discreteness effects due to resolution-limited 
sampling need to be considered, since some continuous approaches may fail~\cite{bray_persistence_2013}. 
Such effects have been investigated for stationary Gaussian Markov processes~\cite{Majumdar_PhysRevE.64.015101,majumdar_spatial_2006} 
and also for some non-Markovian smooth processes~\cite{Ehrhardt_PhysRevE.65.041102}.
A practical analysis of persistence with discrete sampling is discussed for instance in Ref.~\cite{Leeuwen_2009}.


In this paper, we address some finite-size and discretization effects by analyzing interfaces described by a discretized version of Eq.~\eqref{eq:modelrealspace}.  
We analyze the statistics of crossings of an interface relative to its center of mass in interfaces of size $L$ subject to periodic boundary conditions. 
Such crossing points will be called \textit{zeros} of the interface, for short. We will be interested in the density of zeros, and in the intervals $\ell$ that separate two consecutive zeros, both in the steady-state and in the non-stationary relaxation from a flat initial condition. Such an initial condition and the use of periodic boundary conditions ensure translational invariant profiles.
The distribution of intervals can thus be written as $P(\ell;t,L)$, in general. Such distribution can be related to the persistence distribution $Q(\ell)$ (here defined as the  probability that two crossings are separated by a distance larger than $\ell$) by $P(\ell)=Q(\ell)-Q(\ell+1)$. Hence, $P(\ell)$ corresponds to the first-passage distribution. 
Varying $\zeta$ in Eq.~\eqref{eq:modelrealspace}
from $\zeta=-1/2$ to $\zeta=3/2$ allows 
us to describe macroscopically flat to super-rough interfaces and observe the relative relevance of discretization and finite effects.
Within the range of values taken by $\zeta$, three regimes can be identified. Such regimes are characterized in terms of the persistence properties, and by their relation to other observables such as the width of the interface. Hence, this work generalizes the study of Ref.~\cite{zamorategui16distribution}.
In particular, we find that for $\zeta<1$ many of the exact 
results obtained for generalized random walks can be applied to partially describe our findings, while it is not the case for $\zeta>1$ where the finite size of the system governs the scaling laws. 
Further, temporal dependences of all the analyzed quantities are well captured by the dynamical length $L_{\text{dyn}} \sim t^{1/z}$. Therefore, a proper understanding and precise characterization of the self-affine steady-state is fundamental.

The paper is organized as follows. In Section~\ref{sec:model}, we present the discrete dynamical model and define the observables of interest.
Section~\ref{sec:stationary} is devoted to the study of rough interfaces in the stationary state. We investigate the distribution of intervals between zeros and their first two cumulants as a function of the roughness exponent $\zeta$ and the size of the interface $L$. We compare the mean value of the interval length with the width of the interface, which can be analytically computed, for the three regimes identified. 
We also study the distribution and first moments of the total number of zeros in the interface as a function of $\zeta$ and $L$ and relate them with the results obtained for individual intervals. In Section~\ref{sec:nonstationary}, we discuss the non-stationary dynamics by analyzing the scaling of the density of zeros as a function of time and system size $L$ for the three regimes of roughness. Additionally, we explore the non-stationary behavior of the interface width.  In Section~\ref{sec:discussions}, we summarize our results and discuss their relevance for experiments. In the Appendices, we discuss some effective models and detail the calculations.

\section{Model and Methods}
\label{sec:model}

In order to solve the fractional dynamics described by Eq.~\eqref{eq:modelrealspace} we consider one-dimensional interfaces of size $L$ described by the continuous height field $u_x(t)$. The interface is discretized along the spatial direction, such that $x=0,1,...,L-1$. Periodic boundary conditions imposing $u_{L}(t) \equiv u_0(t)$ are considered. 
Each $u_x(t)$ can be thought as the displacement of a particle coupled to other particles $u_{x'}(t)$ through the fractional Laplacian. 
The height of the interface is measured relatively to the center of mass of this particle system so that $\sum_{x=0}^{L-1} u_x(t)=0$, which fixes to zero the total area under the interface.

In Fourier space, the discretized dynamics of Eq.~\eqref{eq:modelrealspace}  is
\begin{equation}
\label{eq:continuousEq}
\partial_t u_q(t)=-|q|^z u_q(t)+\hat \eta_q(t),
\end{equation}
where $u_q(t)=\int dx e^{-iqx}u_x(t)/\sqrt{L}$, and $q=2\pi k/L$ with $k=1,...,L-1$. The Fourier-transformed noise $\hat \eta_q(t)$ is Gaussian with $\langle \hat \eta_q(t) \rangle=0$ and $\langle \hat \eta_q(t)\hat \eta_{q'}(t') \rangle=2T\delta_{qq'}\delta(t-t')$. 
The zero area constraint implies that $u_{q=0}(t)=0$ for all times.

From Eq.~\eqref{eq:continuousEq} we can compute analytically some noise averaged quantities.
Starting from a flat configuration ($u_{q}(t=0)=0$ for all $q$), the averaged stationary state of the structure factor is
\begin{equation}
\label{eq:structure}
S_q(t)=\langle |u_q(t)|^2\rangle=Tq^{-z}(1-e^{-2|q|^zt}).
\end{equation}
Equation \eqref{eq:structure} shows that the system relaxes to equilibrium in a typical time $L^z$, with dynamical exponent $z$, towards a self-affine geometry with $S_q = T q^{-(1+2\zeta)}$.
In this model, the roughness exponent is thus related to the dynamical exponent by 
\begin{equation}
\zeta = (z-1)/2.
\end{equation}
For general self-affine interfaces, such relation does not hold. However, if a generic dynamic scaling is satisfied, we expect a more general form $S_q(t)=\langle |u_q(t)|^2\rangle \sim q^{-(1+2\zeta)}(1-e^{-2|q|^zt})$ for small enough $q$ and long enough times. From such expression, we can define the dynamical length $L_{\text{dyn}}(t) \sim t^{1/z}$ such that length-scales smaller that $L_{\text{dyn}}(t)$ get equilibrated at time $t$, while larger length-scales still keep memory of the initial condition.
Likewise, the structure factor $S_q(t)$ appears in the expression of the width 
$w^2(t)\equiv L^{-1}\sum_{x} 
\langle u_x^2(t)\rangle =L^{-1}\sum_{q\neq 0} S_q(t)$ of the 
interface. A precise 
expression for $w^2(t)$ can be obtained from
\begin{equation}
w^2(t) \approx \frac{T}{\pi} 
\int_{2\pi/L}^{\pi} dq\;|q|^{-z} [1-\exp(-2|q|^z t)].
\label{eq:wtfromSq}
\end{equation}

From this expression, it is easy to show that 
$w^2(t)\sim 2T t$ for very short times. 
At larger times and $\zeta>0$, we get
$w^2(t)\sim L_{\text{dyn}}(t)^{2\zeta}$ for intermediate times,  
and $w_s^2 = w^2(t\to \infty) \sim L^{2\zeta}$ in the steady state limit. See the Appendix~\ref{app:widthanalytic} for details.

Unfortunately, none of the above quantities give us access to the statistical properties of the zeros of $u_x(t)$. 
Indeed, by writing $u_q(t) = |u_q(t)| e^{-i\phi_q(t)}$, we notice that the zeros are 
particularly sensitive to the relative phases 
$\phi_q(t)$ of the modes (see Appendix \ref{app:low-modes} for an illustrative example). 
Therefore, the statistical properties of zeros, even for the simple 
linear model of Eq.~\eqref{eq:modelrealspace}, are highly non-trivial.

In order to study the zeros of $u_x(t)$ as a function of time we will first solve iteratively the dynamics in Fourier space 
with a time-discrete version of Eq.~\eqref{eq:continuousEq},
\begin{equation}
 \label{discrete}
 u_q(t+\Delta t)= \frac{(1-\tfrac{\Delta t}{2}\Lapq)u_q(t)+\sqrt{TL\Delta t}\eta_q(t)}{1+\tfrac{\Delta t}{2}\Lapq}
\end{equation}
with $u_q(t)=\sum_x e^{-iqx}u_x(t)/\sqrt{L}$ and $q=2\pi k/L$.  $\Lapq$ is the exact Laplacian in Fourier space given by $\Lapq=[2(1-\cos q)]^{z/2}$. 
We denote by $\Delta t$ the time step. 
In this paper, we only consider the 
flat initial condition $u_{q}(t=0)=0$.
The choice of the Stratonovich discretization~\cite{stratonovich_topics_1967} in Eq.~\eqref{discrete} is the most appropriate representation, as discussed in a previous work~\cite{zamorategui16distribution}. In this paper, a general condition on the time-discretization $\Delta t$ was derived, as a generalization to the well-known Von Neumann stability criterion~\cite{press_numerical_2007}. The discrete noise is generated by sampling a time sequence of $L$ uncorrelated Gaussian random numbers.
From $u_q(t)$, we can get $u_x(t)$ to obtain the zeros of the interface, and perform a statistical analysis by sampling many noise histories. 
The whole scheme just described can be implemented very efficiently for large interfaces using graphics processing units. 
To do so we exploit the parallelism of the dynamical evolution in Fourier space and
use a parallel random number generator to generate $\eta_q(t)$. In order to get $u_x(t)$, we use parallel fast Fourier (anti)transforms. The detection of zeros in $u_x(t)$ can also be implemented efficiently using parallel search algorithms.

To study the stationary state we optimize the scheme described above by directly sampling independent configurations from the equilibrium Boltzmann distribution 
${\cal P}[u_q] \propto \exp[-\frac{1}{T} \sum_{k=0}^L |q|^z |u_q|^2]$.
This is achieved by generating the complex modes amplitudes
\begin{equation}
\label{eq:amplitude}
 \Re(u_q)\distr\Im(u_q)=\sqrt{\frac{T}{2q^{z}}}\eta_q
\end{equation}
where $\distr$ indicates ``equal in distribution'', which verifies $S_q = \langle |u_q|^2 \rangle = T q^{-z}$ as desired in the steady-state. Consequently, the steady state can be then compared with the long-time limit of the non-stationary relaxation.

The position of a \textit{zero} of the discretized interface is given by the immediate integer to the left of the crossing point at which the interface height changes its sign. This definition was shown~\cite{zamorategui16distribution} to describe correctly the first-passage distribution of the interface in the case $\zeta=1/2$. In the steady-state, we will be interested in the distribution $\pofl$ of the length $\ell$ of the intervals between consecutive zeros, and also in the total number of zeros $n$ and its distribution $P(n)$. We will compare the first moments of these distributions with the mean steady-state width 
\begin{equation}
w_s \equiv \lim_{t\to \infty}w(t) 
\label{eq:ws}
\end{equation}
where 
\begin{equation}
w(t)=\sqrt{\sum_x \langle u_x(t)^2/L \rangle}.
\label{eq:wt}
\end{equation}
In the non-stationary state we will focus in the mean density of zeros 
\begin{equation}
\langle\rho(t)\rangle=\langle n(t) \rangle/L,
\label{eq:rhot}
\end{equation}
as a function of time and compare it with the non-stationary width of the interface $w(t)$.    

\begin{figure}[h!]
 \centering
\includegraphics[scale=0.75]{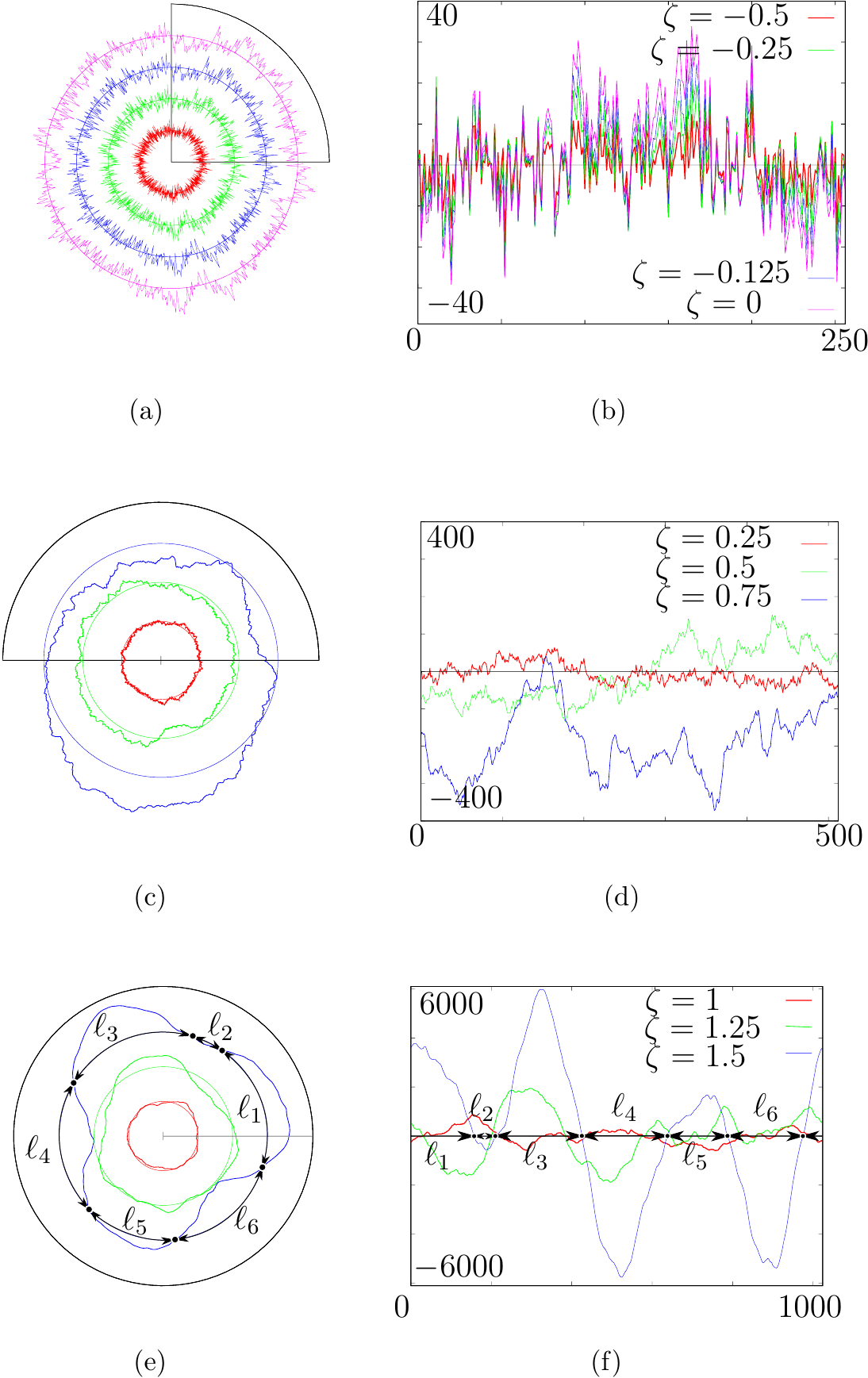}
 \caption{Steady-state configurations with different roughness exponent $\zeta\in[-1/2,3/2]$ in a lattice of size $L=1024$ with periodic boundary conditions (PBCs). Figures on the left show the whole configurations on a circle, illustrating the PBCs with the radius of the circle chosen arbitrarily for presentation purposes. The quadrant of the circle in black corresponds to the segment of the interface depicted in the figures on the right. The interfaces on the right have their center of mass around zero which allows us to appreciate their relative amplitude. (a), (b) Typical interfaces with roughness exponent $\zeta\leq 0$ and the same initial noise. (c), (d) Typical configurations in the interval $\zeta\in(0,1)$. (e), (f)  Rare interfaces with $\zeta\geq1$ for which small intervals are observed. For $\zeta=3/2$ we show the intervals $\ell$ and the zeros of the configuration. Both the PBCs and the fact that the interface vibrates around its center of mass determine the distribution of the intervals. The amplitude of each mode in the Fourier series decay as $\sqrt{\frac{TL}{2q^{z}}}$ with $q=2\pi k/L$ (see Eq.~\eqref{eq:amplitude}). Therefore, the larger the value of $\zeta$, the more relevant the low-frequency modes are. \label{configurations}}
\end{figure}

\section{\label{sec:stationary} Stationary state}

We devote this section to the study of the steady-state limit. 
In Fig.~\ref{configurations} we show typical configurations sampled according to 
Eq.~\eqref{eq:amplitude} for some values of $\zeta$ in regimes \RN{1} ($\zeta< 0$), regime \RN{2}  ($0<\zeta<1$) and regime \RN{3} ($\zeta>1$). 
The center of mass of the interfaces is fixed to zero in all cases.
We can observe that large values of $\zeta$ produce large excursions as the amplitude of the Fourier modes decays as $\sim 1/q^{1 + 2\zeta} \sim L^{1+2\zeta}$. For negative $\zeta$, all the modes have a non-divergent amplitudes in the thermodynamic limit and consequently excursions are typically small. As we already mentioned, boundary conditions and finite-size effects may play an important role constraining long excursions through the zero area condition.

\subsection{Interval distributions}
\label{sec:first-passage}

We first analyze the intervals $\ell_i$ separating the $n$ consecutive zeros of a configuration, as shown in Fig.~\ref{configurations}. 
In general we observe that for $\zeta<0$ the number $n$ of zeros is large and intervals are small compared with the interface size $L$. On the contrary, for large positive $\zeta$ the number of zeros is small as excursions are large. In this regime several intervals are of order $L$. Interestingly, even for large $\zeta$ we can still observe small intervals coexisting with very large ones.  

In order to quantify interval fluctuations we compute the probability distribution as
\begin{equation}
P(\ell)=\left \langle \sum_{i=1}^{n} \delta_{\ell, \ell_i} \right\rangle.  
\label{eq:pofl}
\end{equation} 
The random variable $\ell$ is thus ``local'' and contributes $n$ times to $P(\ell)$ in a single configuration, as opposed to the random variables $n$, $|u_q|^2$ or $\sum u_x^2/L$ which are global random variables of each configuration.
In~\cite{zamorategui16distribution}, we studied in detail the one dimensional case with $\zeta=1/2$ corresponding to the one-dimensional Edwards-Wilkinson equation. We showed that a truncated form of the Sparre-Andersen theorem~\cite{andersen_fluctuations_1953}, which remains valid for describing the interval distribution, is found to be $\pofl \sim \ell^{-3/2}$ below a size dependent cut-off. In spite of the cut-off, the power-law exponent can be related to the persistence exponent $\theta=1/2$ of (infinite-size) Brownian motion as $3/2 = 1+\theta$. In this paper we will extend such study to the range~$\zeta\in[-1/2,3/2]$.

\begin{figure}[h!]
 \centering
\includegraphics[scale=0.9]{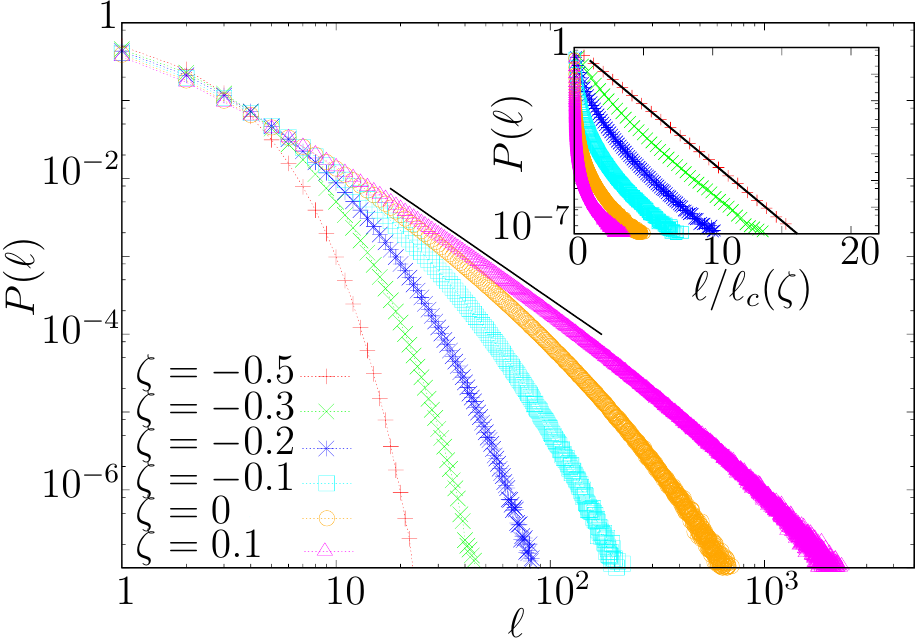}
 \caption{First-passage distribution $P(\ell)$ for $L=131072$ with $\zeta\in[-1/2,0.1]$ corresponding to the regime~\RN{1} and the transition towards the regime~\RN{2}. For $\zeta=-1/2$, the distribution of the intervals is a pure exponential $\pofl=(1/2)^{\ell}=\exp(-\ell \log(2))$. As $\zeta$ increases, correlations among intervals develop and the distribution of the intervals is exponential $P(\ell)\sim \exp(-\ell/\ell_c(\zeta))$ only for intervals $\ell>\ell_c(\zeta)$. The value $\ell_c(\zeta)$  is fitted from the corresponding exponential regime for each $\zeta$. For $\zeta>0$ but close to zero, a power law $P(\ell)\sim \ell^{-\gamma}$ followed by an exponential cut-off is observed. The solid line shows the power-law behavior in the $P(\ell)$ with $\gamma=2-\zeta=1.9$. The inset shows $P(\ell)$ as a function of $\ell/\ell_c(\zeta)$. The solid line corresponds to the function $f(x)=e^{-x}$ that describes exactly the distribution $P(\ell)$ expected for the Ornstein--Uhlenbeck process. We observe that for all $\zeta<0$, an exponential regime is observed for $\ell>\ell_c(\zeta)$. 
\label{intervals_smallZetas}}
\end{figure}

The simplest extreme case is $\zeta=-1/2$, when 
Eq.~\eqref{eq:continuousEq} reduces, for the discrete interface, 
to $L$ independent Ornstein--Uhlenbeck processes~\cite{uhlembeck1930brownian}. 
In real space, the zero area condition 
$\sum_x u_x = 0$ implies that particles 
have a mean-field effective interaction fixing the center
of mass. However, for a large system they 
can still be considered as $L$ independent 
Ornstein--Uhlenbeck processes as the center 
of mass vanishes as $1/\sqrt{L}$ when $z=0$.
Therefore, the steady-state probability for $u_x$ to be above or below zero at any point in space is $1/2$, independently of the height of the interface at any other position. Consequently, the probability of having an interval of length~$\ell$ is
\begin{equation}
 \label{pdfzetaneg}
 \pofl=(1/2)^{\ell}=\exp(-\ell \log(2)),
\end{equation}
where we define a characteristic length $\ell_c(\zeta=-1/2)=1/\log 2$. Fig.~\ref{intervals_smallZetas} confirms this exact exponential decay. In the figure we also show that for all $-1/2 \leq \zeta<0$ we still find an exponential 
decay at large enough $\ell$, with $\ell_c(\zeta)$ increasing with increasing $\zeta$. The exponential decay can be associated with the existence of a finite correlation length and the growth of $\ell_c(\zeta)$ can be attributed to the increase of correlations with increasing $\zeta$.

\begin{figure}[h!]
 \centering
 \includegraphics[scale=1.1]{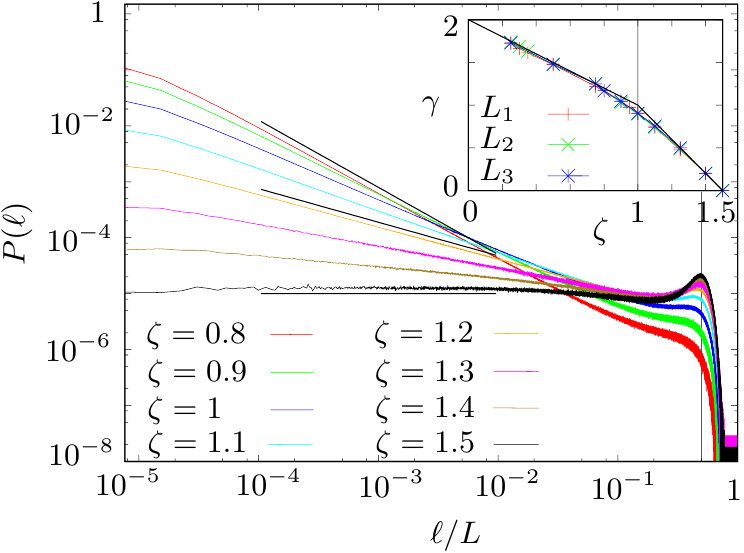}
 \caption{First-passage distribution $P(\ell)$ for $\zeta>0$ as a function of $\ell/L$ for $L=131072$. The vertical line corresponds to the value $\ell/L=1/2$. The distributions behaves as a power law $P(\ell)\sim \ell^{-\gamma}$ for $\ell\gg 1$ but much smaller than the system size. For the regime~\RN{2}, corresponding to $\zeta\in(0,1)$, the exponent is $\gamma=2-\zeta$. For the regime~\RN{3}, corresponding to $\zeta\in(1,3/2)$, the exponent is well described by $\gamma=3-2\zeta$ which was found heuristically (dashed lines). The inset shows the comparison between the power-law exponent $\gamma$ as a function of $\zeta$ given in the previous expressions, and the numerical values measured from the  distributions $P(\ell)$ for $L_1=16384$, $L_2=65536$ and $L_3=131072$ . \label{intervals_largeZetas}}
\end{figure}

Interestingly, as $\zeta=0$ is approached, a power-law behavior emerges in the small $\ell$ regime before the crossover to the exponential decay. Further, as we increase $\zeta$, one observes that $\ell_c(\zeta\approx 0) \sim {\cal O}(L)$ and the cut-off of the distribution localizes at $\sim L/2$. Such cut-off is indeed expected in the limit of very large $\zeta$ where the first modes dominate the fluctuations of the interface (see Appendix \ref{app:low-modes} for an illustrative example).    
As shown in Fig.~\ref{intervals_largeZetas}, for positive values of $\zeta$ below 
the finite-size cut-off, the distribution $\pofl$ is well described by $P(\ell)\sim \ell^{-\gamma}$.
In this regime, the exponent $\gamma$ is expected to be related to the spatial persistence exponent $\theta$ as $\gamma=\theta+1$.
For $0<\zeta<1$ we find $\gamma \approx 2-\zeta$, in good agreement with the prediction $\theta=1-\zeta$~\cite{majumdar_spatial_2001}, made for infinite interfaces. For $\zeta>1$, we find instead 
a crossover towards an unexpected dependence 
$\gamma \approx 3-2\zeta$ (see the inset of Fig.~\ref{intervals_largeZetas}), which does not match the prediction $\theta=0$ for $\zeta>1$. We argue that this discrepancy is ultimately due to the zero area constraint that becomes particularly relevant for \emph{finite} super-rough interfaces. In particular, as we show in Section \ref{ref:nofzeros},  the number of zeros is extensive for $\zeta<0$, subextensive for $0<\zeta<1$, and intensive for $\zeta>1$. 
It is worth stressing, however, that the relation $\gamma \approx 3-2\zeta$ is system-size independent whenever the system is finite, \emph{i.e.}~no matter how large the system is, the zero-area constraint will introduce strong correlations between the intensive number of intervals in the super-rough regime~\footnote{All of our results are for periodic boundary conditions only. We do not investigate how the zero-area constraint modifies the interval distribution in finite systems with different boundary conditions and whether it changes $\gamma(\zeta)$.}.
This observation does not contradict the prediction $\theta=0$~\cite{majumdar_spatial_2001} which is obtained for strictly infinite interfaces. 

It is also worth mentioning that for all $\zeta>0$, we observe  correlations between intervals (see Appendix \ref{app:correlations}).
These correlations were also present in the particular case $\zeta=1/2$ studied in~\cite{zamorategui16distribution}, where successive increments were also shown to be long-range correlated in spite of the local character of the regular Laplacian. These correlations are due to the periodic boundary conditions, since the sum of all the increments that generate the interface add to zero. This constraint is reflected in the correlator of the noise whose off-diagonal elements are all equal and positive in the case $\zeta=1/2$~\cite{zamorategui16distribution}.  Yet the correlator of the increments can be computed for any roughness coefficient $\zeta$, it is beyond the goal of the present article to discuss its exact shape. Regarding the correlation of intervals, we believe that they are strongly related to the zero-area constraint imposed by both the boundary conditions and the fixed center of mass along which intervals are produced (See Appendix \ref{app:correlations}). %


Surprisingly, for the extreme case $\zeta=3/2$ we observe that the distribution of intervals is flat up to a cut-off (see Fig.~\ref{intervals_largeZetas}). By looking at the steady-state configurations, one realizes that although a 
typical configuration has no more than a few zeros, small intervals are rare but still present and the shape of  $\pofl$ shows that they are equally likely.
In Appendix \ref{app:low-modes}, we investigate this behavior in detail by gradually increasing the number of Fourier harmonics in a random Gaussian signal. 
In particular, for $\zeta=3/2$ we find that the suppression of short wavelength modes induces a linear behavior $\pofl \sim \ell$ for intervals smaller than the minimum cut-off wavelength. Next, a saturation to a flat distribution $\pofl \sim \ell^0/L$ for larger intervals  is observed, before reaching a ``hump'' centered around $\ell=L/2$ which contains the most probable intervals.
In Fig.~\ref{intervals_largeZetas}, we see that in general such a hump 
smoothly develops for $\zeta>0$ and becomes more pronounced 
as we increase~$\zeta$. 
The presence of the hump is a sign of the importance of large intervals of length $\ell\sim O(L)$. This fact is understood by comparing the relative typical amplitude of the Fourier modes. If we compare the typical amplitude of the $n_1$ and $n_2>n_1$ harmonic we have indeed $S_{q_1=2\pi n_1/L}/S_{q_2=2\pi n_2/L} = (n_2/n_1)^{1+2\zeta}$ 
which becomes large for large $\zeta$. 
In particular, for the fundamental mode $n_1=1$ and the first harmonic we obtain for $\zeta=3/2$ that $S_{q_1}/S_{q_2}= 16$.
While the fundamental mode has trivially two zeros and $\pofl \sim \delta(\ell - L/2)$, 
for a Gaussian random signal that combines the first two modes  
we observe that the hump is broadened but remains peaked around $L/2$ (see discussion in Appendix \ref{app:low-modes}). 
While we find that the hump is already visible for $\zeta \gtrsim 1/3$, 
the hump dominates the interval statistics for $\zeta>1$, as explained below.

\begin{figure}[t!]
 \centering
     \subfigure[][]{\includegraphics[scale=1]{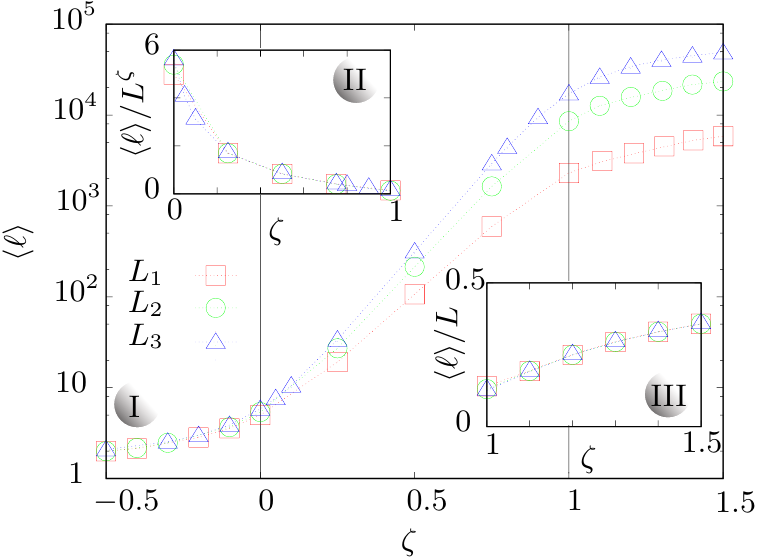}\label{meanl}}
  \subfigure[][]{\includegraphics[scale=1]{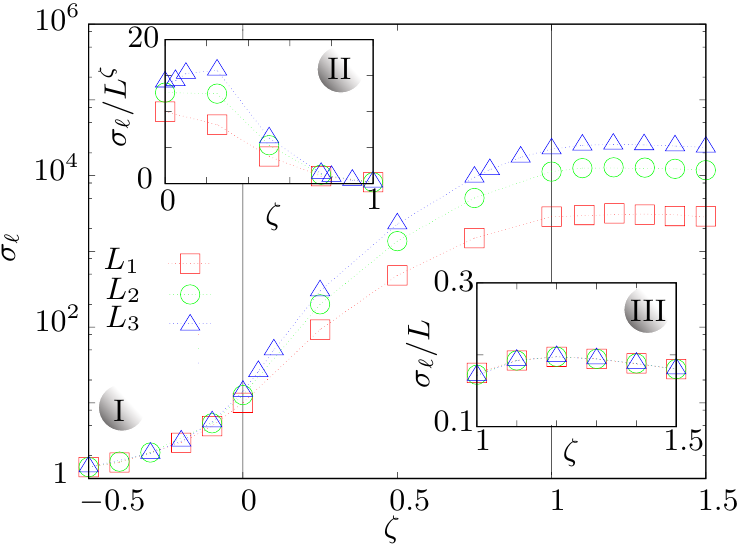}\label{variancel}}
\caption{\subref{meanl} Mean value of the interval $\langle \ell \rangle$ as a function of $\zeta$. In general we identify three regimes for $\zeta$ that scale differently with the size of the system $L$. Each regime is depicted by roman numerals. For Regime~\RN{1}, the mean value $\langle \ell \rangle$ is independent of the size of the system. The insets show the collapse of all the values $\langle \ell \rangle$ for the different system sizes for $\zeta>0$: Regime~\RN{2}, $\zeta\in(0,1)$, where $\langle \ell\rangle\sim L^{\zeta}$. The values depicted in the ordinates correspond to the proportionality factor. Regime~\RN{3}, $\zeta>1$, where $\langle \ell \rangle$ reaches the size of the system and  scales as $\sim L$ (the placing of the roman numerals coincide with the regime they represent in the main figure).  The scaling in regime~\RN{3} is understood by the peak displayed by the distribution $\pofl$ around $L/2$, where larger intervals become more probable. \subref{variancel} Standard deviation $\sigma_{\ell}=\sqrt{\langle \ell^2\rangle-\langle \ell \rangle^2}$  computed directly from  the numerical distributions $\pofl$ as a function of the roughness exponent $\zeta$. The inset with the numeral~\RN{2} shows a rescaling of all the values for the standard deviation in the interval $\zeta\in[0,1]$, illustrating that a naive scaling $\sigma_{\ell}\sim L^{\zeta}$ does not hold except for $\zeta=1$. The inset with the numeral~\RN{3} corresponds to the regime for $\zeta>1$ where the scaling is $\sigma_{\ell}\sim L$ . For $\zeta<0$ both moments are independent of the system size $L$. 
\label{moments}}
\end{figure}

In  Fig.~\ref{moments}, we show the first two cumulants $\langle \ell \rangle$ and $\sigma_{\ell}=\sqrt{\langle \ell^2 \rangle-\langle \ell \rangle^2}$ of the distribution $P(\ell)$ in the whole interval of $\zeta$ analyzed. The study of the cumulants as a function of $\zeta$ allows us to clearly identify three regimes: 
For the values of  $\zeta\in[-1/2,0)$, the mean value $\langle \ell \rangle$ and the standard deviation $\sigma_{\ell}=\sqrt{\langle \ell^2 \rangle-\langle \ell \rangle^2}$ do not depend on $L$. This regime is identified in the following as Regime~\RN{1}. For $\langle \ell \rangle$ we expect to have $\langle \ell \rangle\approx \ell_c(\zeta)$ which is related to the decay rate in the exponential distribution shown in Fig.~\ref{intervals_smallZetas}. 
For $\zeta\in(0,1)$, $P(\ell)$ presents the scale-invariant regime $\pofl \sim \ell^{-\gamma} = \ell^{-(2-\zeta)}$, consistent with the persistence exponent  $\theta=1-\zeta$ of the infinite interface. As shown in the top inset of Fig.~\ref{moments}(a) in this regime we 
find $\langle \ell \rangle \sim L^{\zeta}$. This intermediate regime will be identified as Regime~\RN{2}. 
Finally, for the regime corresponding to $\zeta\in(1,3/2]$, small intervals become rare and the mean value is dominated by the large intervals of length of order $\sim L/2$. As shown in the bottom inset of Fig.~\ref{moments}(a) we find $\langle \ell \rangle \sim L$ (with only a weak dependence on $\zeta$ near the crossover at $\zeta=1$) and $\sigma^2_{\ell} \sim L^2$. This regime will be referred to as Regime~\RN{3}, or super-rough regime. 
The behavior of the moments observed numerically can be obtained by focusing on the decay behavior of $\pofl$ simply by introducing an infrared and an ultraviolet cutoffs, $\ell_0 \sim L$ and $\ell_0 \sim 1$, respectively.
While for the regime~\RN{1}, $\langle \ell \rangle$ is $L$ independent, for the regime~\RN{2} and~\RN{3}, we obtain 
\begin{equation}
\langle \ell \rangle
\approx \frac{\int_{l_0}^{L_0}\ell^{-\gamma+1} d\ell}{\int_{l_0}^{L_0} \ell^{-\gamma}  d\ell} 
= \left(\frac{1-\gamma}{2-\gamma}\right) \frac{L_0^{2-\gamma} - l_0^{2-\gamma}}{L_0^{1-\gamma} - l_0^{1-\gamma}}. 
\end{equation}
Thus, in the large size limit we find $\langle \ell \rangle \sim {\cal O}(L^{\zeta})$ for $0<\zeta<1$,  and $\langle \ell \rangle \sim {\cal O}(L)$ for $\zeta>1$. In consequence, the crossover from regime \RN{2} to~\RN{3} at $\zeta=1$ represents a transition from a $\zeta$-dependent $\langle \ell \rangle$ to a $\zeta$-independent behavior. Similar observations can be performed for $\sigma_{\ell}$ as shown in Fig. \ref{moments}(b) in the regime I and III, while the scaling of $\sigma_\ell$ is non-trivial in the regime II due to behavior of $\langle\ell^2\rangle$.

\begin{figure}[h!]
 \centering
 \includegraphics[scale=1]{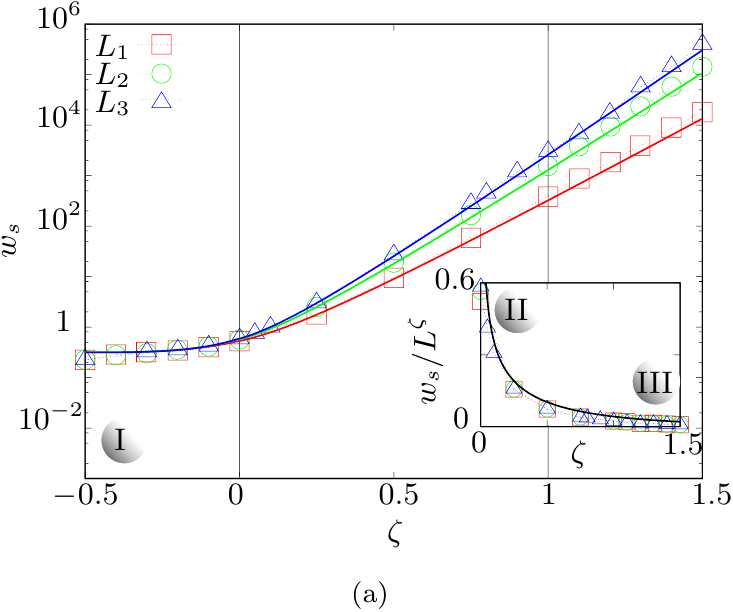}
 \caption{Stationary width $w_s=w(t\to\infty)$  as a function of the roughness coefficient $\zeta$ for $L_1=16384$, $L_2=65536$ and $L_3=131072$. The solid lines represent the long-time behavior of the width given by the Eq.~\eqref{ws2}. For $\zeta<0$,~\emph{i.e.}~regime~\RN{1}, the width $w_s$ is independent of the system size $L$ and is dominated by the high-frequency modes.  The definition of the roughness exponent $\zeta$ comes from the dependence of the width on the system size at the saturation time $w_s\sim L^{\zeta}$ for $\zeta>0$. The width in this case is dominated by the low-frequency modes.  The inset shows the collapse of all the curves for different system sizes $L$ for regimes~\RN{2} and~\RN{3} . In these regimes, the width scales as $\langle w_s \rangle\sim L^{\zeta}$, as shown in Eq.~\eqref{wsregime2and3}. The solid line corresponds to the prefactor $\frac{w_s}{L^{\zeta}}\approx \sqrt{\frac{T}{(2\pi)^{1+2\zeta}\zeta}}$ of Eq.~\eqref{wsregime2and3}. 
 \label{widthvszeta}}
\end{figure}

It is interesting to compare the behavior of $\langle \ell \rangle$ with the stationary width of the interface  $w_s$ defined in Eq.~\eqref{eq:ws}. 
From Eq.~\eqref{eq:wtfromSq} we get, for $\zeta \neq 0$
\begin{equation}
 \label{ws2}
 w_s^2 \approx \frac{T}{2\pi \zeta}\left[\left( \frac{L}{2\pi} \right)^{2\zeta}- \pi^{-2\zeta} \right],
\end{equation}
where we used $z=1+2\zeta$. 
This analytical expression is in perfect agreement with the numerical results, as shown in Fig.~\ref{widthvszeta}. 
The case  $\zeta=0$ is marginal, 
\begin{equation}
w_s^2 \approx 
\frac{T}{\pi} \log(L/2)
\end{equation}

From Eq.~\eqref{ws2}, we can obtain large-$L$ expressions for each regime of $\zeta$ as follows. 
For the regime~\RN{1}, $\zeta<0$, the stationary width $w^2_s$ is dominated exclusively 
by the ultraviolet cut-off :
\begin{equation}
\label{wsregime1}
 w^2_s\approx\frac{T}{2\pi^{1+2\zeta}|\zeta|}.
\end{equation}
This independence in $L$ for $\zeta<0$ (regime~\RN{1}) 
can be appreciated in Fig.~\ref{widthvszeta} when comparing three different sizes. 
If $\zeta=-1/2$, one has $w^2_s=T$ as expected for an Ornstein--Uhlenbeck process, or by 
the energy equipartition theorem. 

For the regimes~\RN{2} and~\RN{3}, the stationary width is dominated by the infrared cut-off 
\begin{equation}
\label{wsregime2and3}
w^2_s\approx \frac{T}{(2\pi)^{1+2\zeta} \zeta} L^{2\zeta} 
\end{equation}
as seen from the numerical results 
in the inset of Fig.~\ref{widthvszeta}, 
which also shows the analytical prefactor.
\begin{figure}[h!]
 \centering
\includegraphics[scale=1]{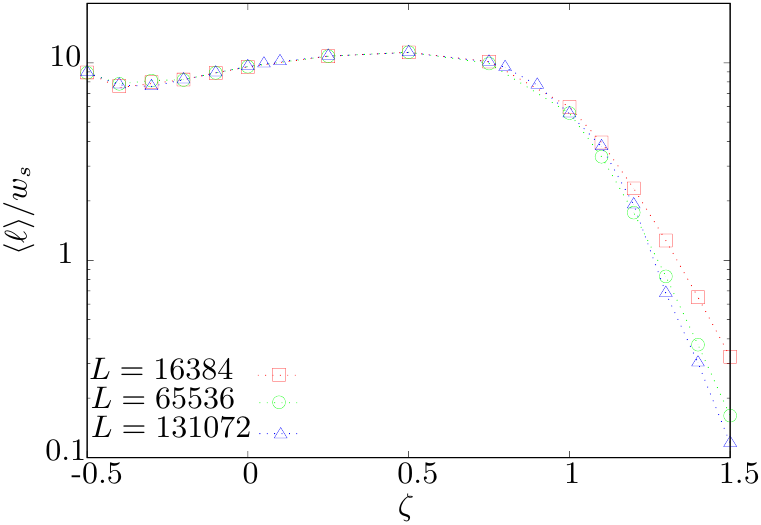}
\caption{Comparison between $\langle l \rangle$ and $w_s$ as a function of roughness exponent $\zeta$ and the system size $L$. }
\label{fig:comparison1}
\end{figure}
Let us now compare $\langle \ell \rangle$ and $\langle w_s \rangle$, shown in 
Fig.~\ref{moments} and~\ref{widthvszeta}. 
We see that $\langle \ell \rangle$ and $w_s$ follow the same  
scaling with $L$ in regime~\RN{1} and~\RN{2}. More remarkably, the scaling prefactor, which depends on 
$\zeta$ is also very close. The average interval, measuring fluctuations in the longitudinal direction,  
is thus controlled by the width of the interface for $-1/2 < \zeta < 1$ (regimes \RN{1} and \RN{2}), 
which measures fluctuations in the transverse direction. 
However, in regime \RN{3} this relation is
broken as $\langle \ell \rangle$ is constrained by the size of the 
system while $w_s$ continues to grow as $L^{\zeta}$. 
In Fig. \ref{fig:comparison1} we show that 
they can be indeed approximately considered to be proportional for all the values $\zeta<1$, as $\langle \ell \rangle \approx 10\, w_s$, neglecting a small $\zeta$ dependent correction of order ${\cal O}(1)$.
This connection between a quantity that measures fluctuations in the transverse direction and one that measures fluctuations on the longitudinal direction is interesting, particularly because $w_s$ can be computed analytically as discussed in the section \ref{app:widthanalytic} of the Appendix.

\subsection{Fluctuations in the total number of zeros}
\label{ref:nofzeros}
The periodic boundary conditions and the zero-area condition fixing the center of mass position impose a non-trivial constraint 
on the Fourier modes complex amplitudes of Eq.~\eqref{eq:amplitude}. 
General considerations can be made about the number of zeros in truncated Fourier series.
In particular, the number of zeros is always even and there is a maximum of $2N$ zeros (if $q_{cut} = 2\pi N/L$ is the shortest wavelength in the series),
and a minimum of $2$ zeros if the zero-area constraint is enforced (see Appendix \ref{app:low-modes}).
An illustrative example is the case of a random signal 
composed by only two modes, the fundamental and the first harmonic, discussed in details 
in Appendix~\ref{app:low-modes}. While the fundamental mode alone has exactly two zeros, mixing it with the first harmonic can produce interfaces with four zeros. Under certain conditions, we can compute the probability to observe either two or four zeros, as shown in Appendix \ref{app:low-modes}. 
The task of estimating the probabilities of the allowed number of zeros by 
mixing more modes however becomes analytically intractable. This fact motivates the numerical study of the distribution of the number of zeros in the steady-state.

\begin{figure}[h!]
 \centering
 \includegraphics[scale=0.65]{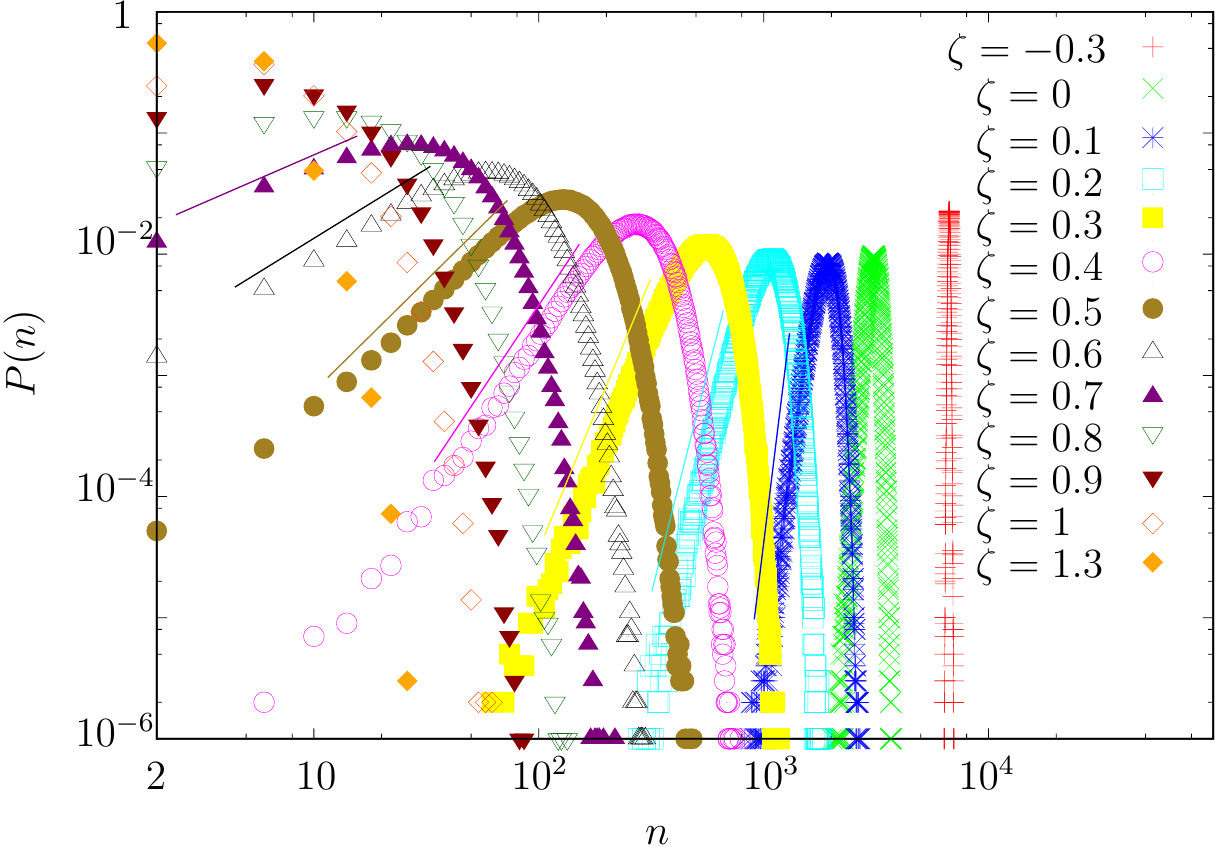}
 \caption{Distributions of the number of zeros $P(n)$ for interfaces of size $L=16384$  for different values of the roughness exponent $\zeta$. As $\zeta$ increases above $\zeta=0$ a power-law regime is observed, as shown by the solid lines. We observe that for large $\zeta$ the distribution is concentrated around $n=2$.  \label{pofnzerosall}}
\end{figure}

Figure~\ref{pofnzerosall} shows the distribution $P(n)$ of the number of zeros for different roughness exponents $\zeta$. 
In the regime~\RN{1}, corresponding to $\zeta<0$, we observe a narrow distribution with a well defined mean value. 
As we gradually increase $\zeta$, we observe that the center of the distribution shifts towards the left, 
while the distributions develops a left tail which behaves approximately as $P(n)\sim n^{2(1-\zeta)/\zeta}$ for values of $\zeta$ around $1/2$, 
as shown by the straight lines in Fig.~\ref{pofnzerosall}. 
Further, for larger $\zeta$ the left tail reaches the minimal value $n=2$, whose probability becomes larger, 
indicating the dominance of the fundamental mode in the super-rough phase. 
%

\begin{figure}[h!]
 \centering
\subfigure[][]{\includegraphics[scale=0.82]{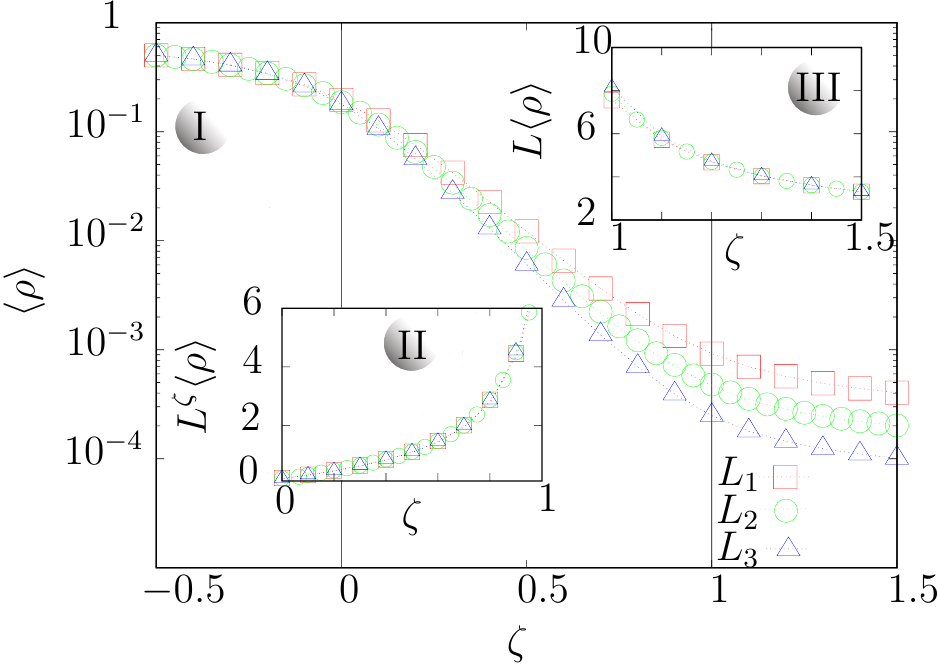}\label{meann}}
\subfigure[][]{\includegraphics[scale=0.82]{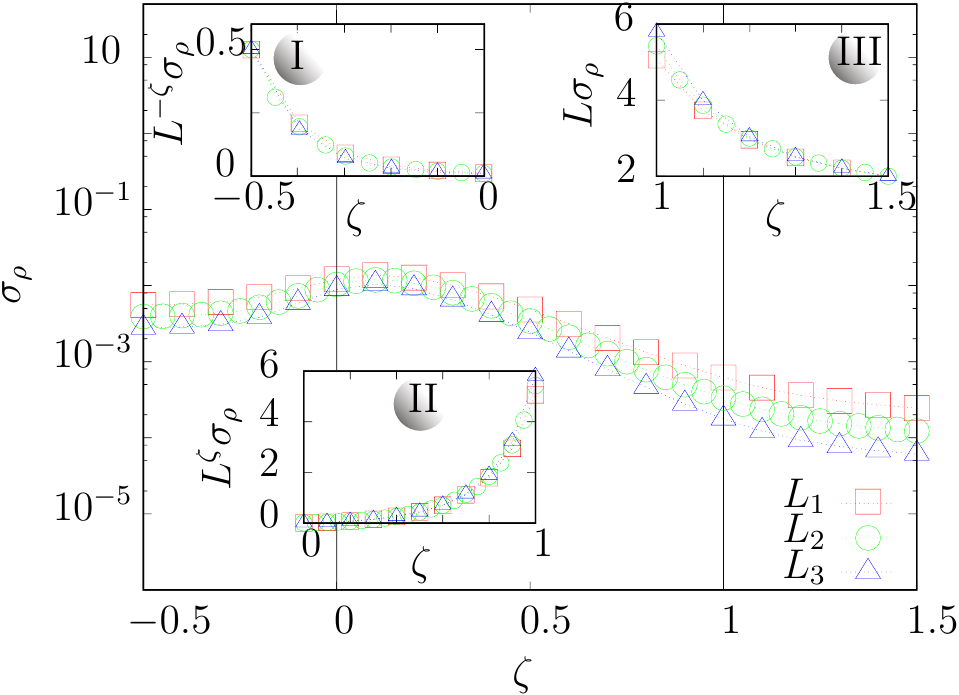}\label{variancen}}
\caption{First two cumulants for the density of zeros $\rho=n/L$, with $n$ the number of zeros, in systems of size $L_1=8192$, $L_2=16384$ and $L_3=32768$. \subref{meann} Stationary mean density $\langle\rho\rangle= \langle n\rangle/L$ as a function of $\zeta$ for $\zeta\in[-1/2,3/2]$. For $\zeta<0$, the number of zeros $\langle n\rangle$ depends linearly on $L$. Hence, the mean density $\langle\rho\rangle$ is independent of $L$. For $\zeta\in(0,1)$, the mean density $\langle\rho\rangle$ depends on the roughness coefficient as $\langle \rho \rangle\sim L^{-\zeta}$.  For $\zeta>1$, the mean number of zeros $\langle n \rangle$ becomes independent of $L$ since typical configurations, dominated by the first two non-trivial modes, have two or four zeros, independently of the system size. Thus, the density behaves as $\langle \rho \rangle\sim L^{-1}$.  \subref{variancen} Standard deviation $\sigma_{\rho}=\sqrt{\langle \rho^2 \rangle-\langle \rho \rangle^2}$ for the density of zeros. In  regimes~\RN{2} and~\RN{3}, the scaling for the standard deviation $\sigma_{\rho}$ corresponds to the one observed for  $\langle \rho \rangle$,~\emph{i.e.}~$\sigma_{\rho}\sim L^{-\zeta}$ and $\sigma_{\rho}\sim L^{-1}$, respectively. The standard deviation $\sigma_{\rho}$ presents a maximum at around $\zeta=0.1$. Interestingly, in the regime~\RN{1} a scaling  $\sigma_{\rho}\sim L^{\zeta}$ is found, which means that in the infinite-size limit the distribution $P(\rho)$ resembles a Dirac delta centered in $\langle \rho \rangle$.  
\label{densityvszeta}}
\end{figure}

\begin{figure}[h!]
 \centering
\includegraphics[scale=1]{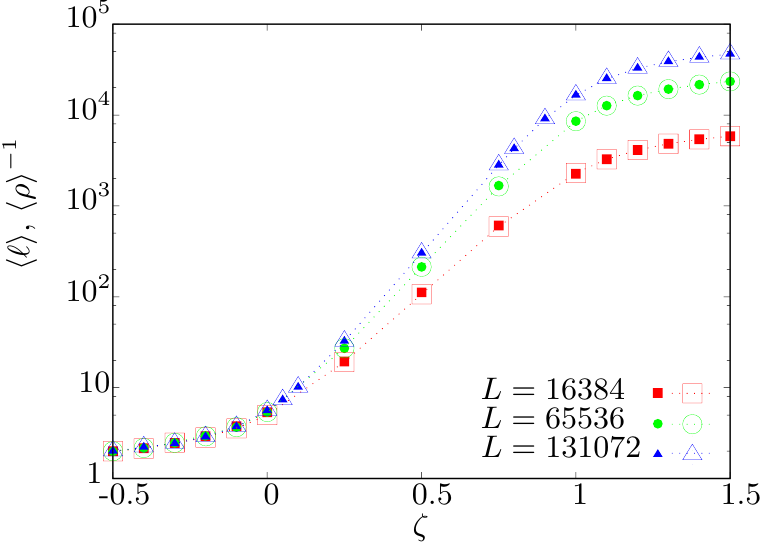}
\caption{Comparison between $\langle \ell \rangle$ (large empty markers) and $\langle \rho \rangle^{-1}$ (small solid markers) as a function of roughness exponent $\zeta$ and system size $L$. }
\label{fig:comparison}
\end{figure}

In order to shed light on the behavior of the average number of zeros $\langle \rho \rangle$, we now compare its value to the average size $\langle \ell \rangle$ of the intervals.
In Fig.~\ref{meann} we show that $\langle\rho\rangle$ displays the same 
properties as $\langle \ell \rangle^{-1}$ in the three regimes, not only the scaling with $L$, but also in the $\zeta$-dependent scaling prefactor, as can be appreciated 
by comparing  with the insets of Fig.~\ref{meanl}. 
%
If a generic scaling $\langle \ell \rangle \approx \langle \rho \rangle^{-1} \equiv L/\langle n \rangle$ is reasonable, there is no reason \emph{a priori} for it to hold exactly.
Indeed, for instance, the number $n$ of zeros is a single random variable for each configuration, while there are precisely $n$ intervals for that configuration contributing to $\langle \ell \rangle$ (see definition of $\pofl$ in Eq.~\eqref{eq:pofl}).
Remarkably, we find however that the behavior of 
$\langle \rho \rangle=\langle n \rangle/L$ and of $\langle \ell \rangle^{-1}$ are equal for all $\zeta$. In Fig \ref{fig:comparison} we show, rather strikingly, that actually $\langle \rho \rangle \approx \langle \ell \rangle^{-1}$, so that even their prefactors are numerically indistinguishable. 
We understand this relation from the fact that the distribution $P(n)$ is sharply peaked (for large $L$) around its most probable value $n^\star$.
%

As we observe in the configurations sketched in Fig.~\ref{configurations}, for $\zeta<0$ the interfaces stay close to their center of mass since the amplitude of the different modes that describe the interface differ slightly from each other. In this regime \RN{1}, 
we find that $n$ is an \textit{extensive} observable, \emph{i.e.}~the number of zeros behaves with the system size as $\langle n \rangle \sim {\cal O}(L)$. This can be appreciated in the size independence displayed by $\langle\rho\rangle=\langle n\rangle/L$ in Fig.~\ref{meann}.
As $\zeta$ increases beyond $\zeta=0$ in regime \RN{2}, larger intervals appear and configurations with fewer number of zeros are more likely to occur. 
We find in this regime a \textit{subextensive} number of zeros $\langle n \rangle \sim {\cal O}(L^{\zeta})$, 
as can be appreciated in the scaling collapse $\langle\rho\rangle L^\zeta$ shown in the left inset of Fig.~\ref{meann}.
The number $n$ of zeros  becomes independent of the system size $L$ at $\zeta=1$, which means that the density of zeros goes to zero in the limit $L\to\infty$, 
or that the mean number of zeros becomes an \textit{intensive} quantity $\langle n \rangle \sim {\cal O}(1)$ in regime \RN{3}. 

Figure~\ref{variancen} shows the standard deviation $\sigma_{\rho}=\sqrt{\langle \rho^2 \rangle-\langle \rho \rangle^2}$, 
which displays a non-monotone behavior as a function of $\zeta$. As for the standard deviation of the intervals, $\sigma_{\rho}$ is controlled by the second moment $\langle \rho^2\rangle$, whose scaling with the system size $L$ is non-trivial. Nonetheless, for the scaling $L^{-\zeta}\sigma_{\rho}$ we observe a perfect collapse for different system sizes.

\section{\label{sec:nonstationary}Non-stationary state}
In this section we consider the non-stationary relaxation of interfaces 
with different $\zeta$ starting from a flat initial condition. 
This case is experimentally relevant since the equilibration time for an 
interface described by the critical dynamics of Eq.~\eqref{eq:modelrealspace} grows 
as $L^{z}$, with $z$ the dynamic exponent (which coincides with the 
Riesz-Feller order of the fractional Laplacian in our model). Aging 
properties are indeed experimentally observable in many systems and
non-stationary persistence properties can be analyzed.

We are interested in particular in the non-stationary density of zeros $\langle\rho(t,L)\rangle$, and in its comparison with the non-stationary width $w(t,L)$. This comparison is directly motivated by the relation found in the steady-state, where (the inverse of) $\langle\rho\rangle$ and $w_s$ display the same scaling for regimes \RN{1} and \RN{2}, relating a ``longitudinal'' to a ``transversal'' property of the interface (see Fig.~\ref{fig:comparison1}). As the initial condition is a flat interface and the very short time dynamics is diffusive, one has $w^2(t) \approx 2Tt$ (see Appendix \ref{app:widthanalytic}) and, at initial times, we expect a large and extensive  number of zeros,~\emph{i.e.}~$n\sim L$. 
We focus in the temporal decay of such excess of initial zeros by monitoring its density
$\langle\rho(t,L)\rangle$.

\begin{figure}[h!]
 \centering
 \includegraphics[scale=0.7]{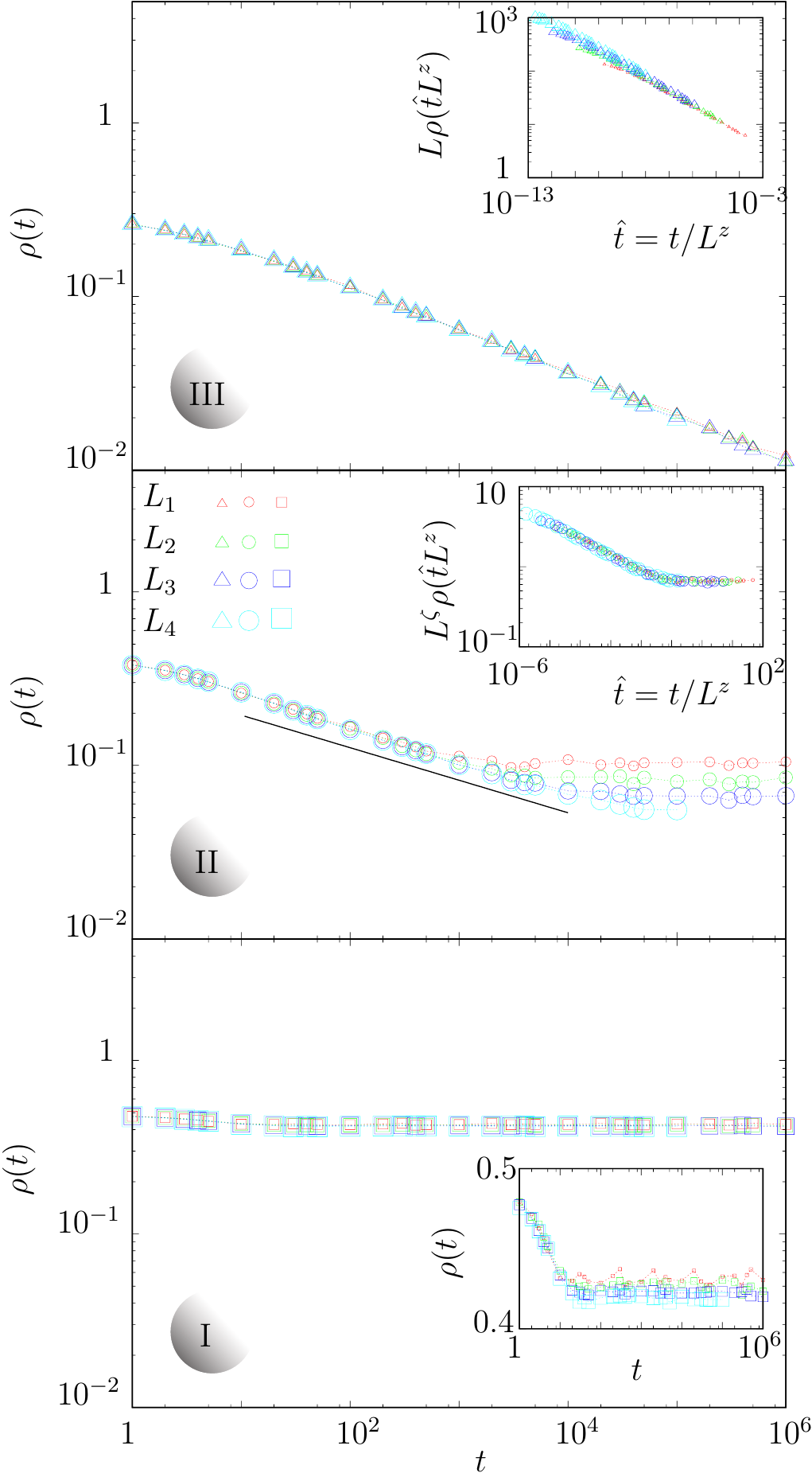}
 \caption{Density of zeros $\langle\rho(t,L)\rangle$ as a function of time $t$ for $L_1=512$, $L_2=1024$, $L_3=2048$ and $L_4=4096$. The values of $\zeta$ shown for each regime are $\zeta=-0.3,0.3$ and $1.3$ for the regimes~\RN{1},~\RN{2} and~\RN{3}, respectively. In the regime~\RN{1}, we do not observe a power-law decay towards the stationary density. Such a decay seems to be system-size independent. For the stationary density given by the plateau, there is a weak dependence on the system size $L$. The stationary case in this regime is approached via an exponential decay. For the regime~\RN{2}, there is a power-law decay of the density of zeros towards the stationary density. The power law is given by the solid line $\sim t^{-\zeta/z}$, with $z=1+2\zeta$. The inset shows the scaled function $\langle\hat{\rho}(\hat{t})\rangle\sim L^{\zeta}\langle\rho(t/L^{z})\rangle$. For the regime~\RN{3}, the saturation time becomes very large $t_s\sim L^{1+2\zeta}$ ($t_s\approx10^9$ for  $L_1=512$), but a scaling function of the form $\langle\hat{\rho}(\hat{t})\rangle\sim L\langle\rho(t/L^{z})\rangle$ is still expected. 
  \label{densityVStime}}
\end{figure}

In Figure~\ref{densityVStime} we show one example of the density of zeros $\langle\rho(t,L)\rangle$ as a function of $t$ for different system sizes $L$ in each regime considered (from the bottom to the top figures they correspond to regimes \RN{1},\RN{2} and \RN{3}, respectively). In all cases, we observe an initial excess of zeros that relaxes towards its stationary value.

For $\zeta<0$, \emph{i.e.}~in the regime~\RN{1}, the steady-state equilibration time is very short and the stationary state is approached via a size-independent exponential convergence, 
shown in Fig.~\ref{densityVStime}.
In this regime the equilibration time does not scale 
with the system size $L$ as expected from dynamic scaling of rough interfaces. 
This is consistent with the existence of a finite correlation length in the steady-state 
of order $\ell_c(\zeta)$, as shown in Fig.~\ref{intervals_smallZetas}.
Thus, the equilibration takes place when the dynamical growing length 
becomes of the order of $\ell_c(\zeta)$.  
In the extreme case for $\zeta=-1/2$, we have a characteristic time
$\tau_c(\zeta=-1/2) \approx 1$ corresponding to the Ornstein--Uhlenbeck process, 
and a monotonic increase in $\tau_c(\zeta)$ with increasing $\zeta$. Such increase in the characteristic time  is consistent with the growth of the correlation length $\ell_c(\zeta)$ and it is ultimately due to the increase of the coupling between neighboring elements of the interface.

For $0<\zeta<1$ in regime \RN{2}, the number of zeros decays as a 
power-law at short times and equilibrates to a size-dependent value.
In the inset of Fig.~\ref{densityVStime} 
we show that it can be fairly described by 
$\langle\rho(t,L)\rangle \sim L^{-\zeta} \langle\hat{\rho}(t/ L^z)\rangle$, so that the equilibration 
time scales as $L^z$, as expected from a simple dynamical scaling. 
Such scaling is consistent with the steady state scaling 
$\langle\rho\rangle \approx \langle \ell \rangle^{-1} \approx w_s^{-1} \sim L^{-\zeta}$.
If we replace $L \to L_{\text{dyn}}(t)=t^{1/z}$, valid for intermediate times 
before equilibration, $L_{\text{dyn}} < L$, we 
find $\langle\rho(t,L)\rangle \sim L_{\text{dyn}}^{-\zeta} = t^{-\zeta/z}$. This behavior is in good agreement with this non-stationary prediction as shown by the solid line in Fig.~\ref{densityVStime}.

In the super-rough regime \RN{3} corresponding to $1<\zeta<3/2$, the number of zeros also 
follows a power-law at short times and equilibrates to a size dependent value, as in regime \RN{2}. Nevertheless, as shown in the inset of Fig.~\ref{densityVStime}, 
the power-law follows a different scaling law $\langle\rho(t,L) \rangle\sim L^{-1} \langle\hat{\rho}(t/ L^z)\rangle$. 
This scaling reflects the fact that in the steady-state, as we go from regime $\RN{2}$ to regime $\RN{3}$, the (inverse of)  the density $\langle\rho\rangle^{-1}$ goes from a $\zeta$-dependent behavior $L^{\zeta}$ to a saturation value $L$. However, the time 
dependence is still controlled by the $\zeta$-dependent 
dynamical length $L_{\text{dyn}}(t)\sim t^{1/z}$, as can be appreciated in the inset where $\langle\rho(t,L)\rangle \sim t^{-1/z}$.

 \begin{figure}[t!]
 \centering
     \includegraphics[scale=0.7]{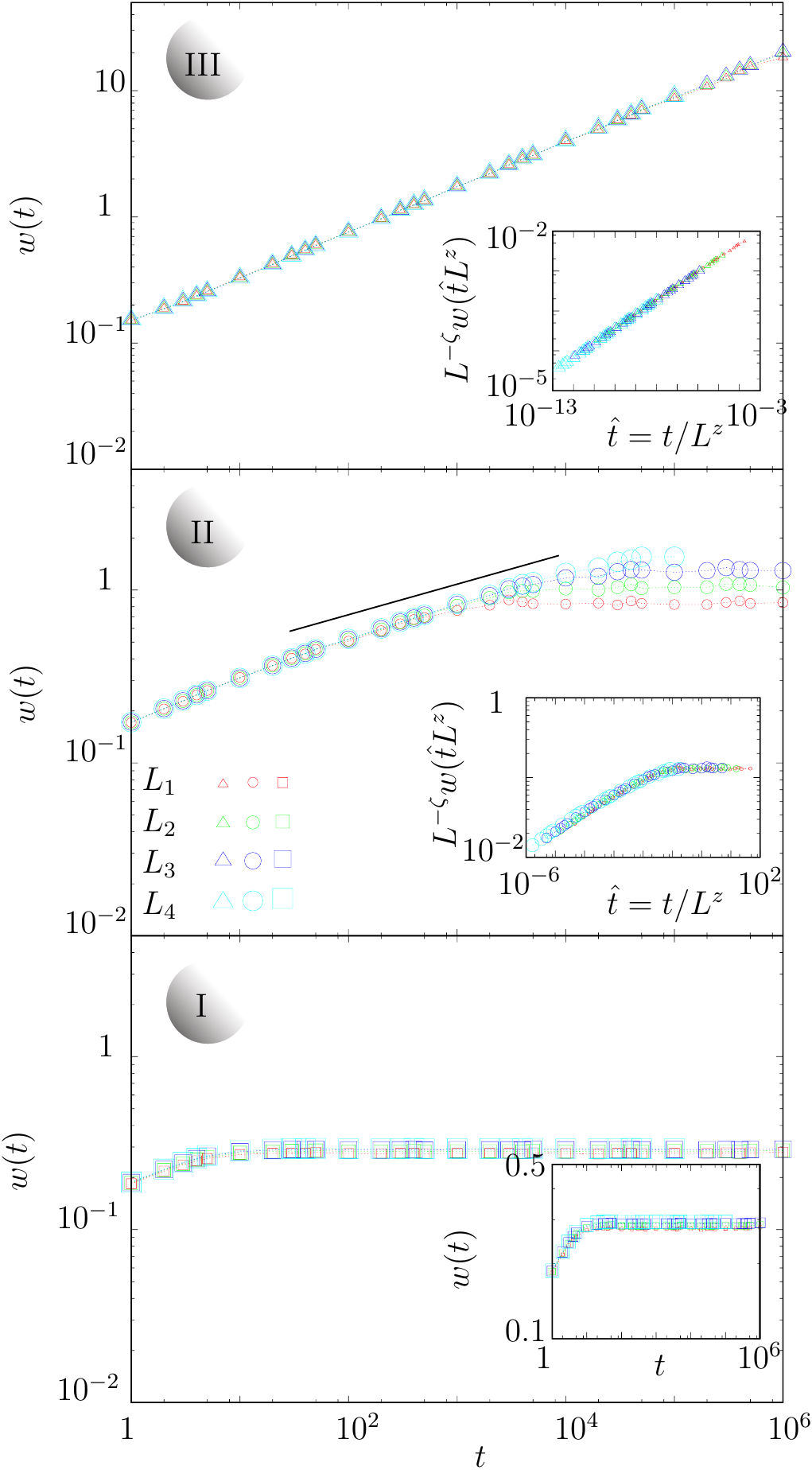}
\caption{Nonstationary width $w(t,L)$ as a function of time $t$ for $L_1=512$, $L_2=1024$, $L_3=2048$ and $L_4=4096$. The values of $\zeta$ shown for each regime are $\zeta=-0.3,0.3$ and $1.3$ for the regimes~\RN{1},~\RN{2} and~\RN{3}, respectively. The initial condition is a flat interface. For the regime~\RN{1}, no scaling is proposed. For the regime~\RN{2}, a power-law behavior $w(t,L)\sim t^{\beta}$ with $\beta=\zeta/z$ describes the evolution of the width for large times, smaller than a saturation time $t<t_s$. We show such power law by the solid lines in black. After a saturation time $t_s\sim L^{z}$ we recover the stationary width $w_s(L)\sim L^{\zeta}$. The inset shows the scaling function for the width is $w_s(t,L)\sim L^{\zeta}\hat{w}(t/L^z)$, where $\hat{w}(\hat{t})$ corresponds to the scaled function independent of $L$ and adimensional time $\hat{t}=t/L^z$. Such scaling is also valid for the regime~\RN{3}. \label{widthVStime} }  
\end{figure}

It is interesting to compare the decay of the number of zeros with the temporal 
increase of the interface width $w(t,L)$, which can be computed analytically (see Appendix \ref{app:widthanalytic}): 
\begin{eqnarray}
w^2(t) 
&\approx& 
t^{2\zeta/z} [F( \pi t^{1/z},z)-F(2\pi L^{-1} t^{1/z},z)],
\label{eq:finitetimes0}
\end{eqnarray}
where 
\begin{eqnarray}
F(y,z)\equiv 
\frac{(e^{-2 y^z}-1) y^{1-z} - 2^{\frac{z-1}{z}} \Gamma[z^{-1}, 2 y^z]}{z-1}, 
\end{eqnarray}
and $\Gamma[a,x]$ is the ``upper'' incomplete gamma function.
In Fig.~\ref{widthVStime} we show the non-stationary evolution of this property for the three different regimes. 
In regime \RN{1}, we observe a size-independent exponentially fast equilibration in the same time scale described by $\rho(t,L)$. 
Accordingly, in regime \RN{2} we observe that $w(t,L)\sim L^{\zeta} {\hat w}(t/L^z )$, showing a tight connection between $w(t,L)$ and $\langle\rho(t,L)\rangle^{-1}$. In contrast, 
in regime \RN{3} we see that $w(t,L)\sim L^{\zeta} {\hat w}(t/L^z)$ 
compared to the scaling $\langle\rho(t,L)\rangle\sim L^{-1} \langle\hat{\rho}(t/L^z)\rangle$ found in the same regime.
This is due to the fact that while the width of the interface is 
not constrained, the number of zeros cannot be smaller than two due to the periodic 
boundary conditions.

In summary, we find that the non-steady relaxation of the mean number of zeros 
in the three roughness regimes can be inferred from the corresponding 
steady-state density of zeros by replacing $L$ by $L_{\text{dyn}}(t)\sim t^{1/z}$ whenever $L_{\text{dyn}}(t)< L$. In such regime we conjecture that a similar rule holds for other quantities, such as the non-stationary interval distribution, $P(\ell;t,L) \sim P(\ell; t=\infty, L_{\text{dyn}}(t)<L)$. 

\section{\label{sec:discussions}Discussion}
%
We have obtained scaling relations for several properties of rough interfaces with periodic boundary conditions both in the stationary and non-stationary state. In the former, properties such as the interval length $\ell$, the  density $\rho$ of crossing zeros or the width $w_s$ of the interface allowed us to identify three regimes. By looking at the first two cumulants of such properties, along with the stationary width, we extracted scaling functions with the system size $L$. Analogously, in the non-stationary state, we identified dynamical scaling functions for both the density of zeros $\langle\rho(t,L)\rangle$ and the width $w(t,L)$ which can be computed analytically. The main scaling results are included in Table \ref{table}.
\begin{table}[h!]
\centering

\begin{tabular}{r|ccc}
 & \RN{1}: $\zeta\in[-1/2,0)$        & \RN{2}: $\zeta\in (0,1)$ & \RN{3}: $\zeta\in(1,3/2]$\\ \hline
$\langle l \rangle \sim$  & $\ell_c(\zeta)$ & $L^{\zeta}$ & $L$\\
$\langle \rho \rangle \sim$               & $\ell^{-1}_c(\zeta)$ & $L^{-\zeta}$  & $L^{-1}$\\ 
$\langle \rho(t,L) \rangle =$               & $\langle\rho(\zeta)\rangle (1-e^{-t/\tau_c(\zeta)})$ & $L^{-\zeta}\langle\hat\rho(t/L^z)\rangle$ & $L^{-1}\langle\hat{\rho}(t/L^z)\rangle$\\
 $w_s \sim$               & ${w_c(\zeta)}$ & ${w(\zeta)} L^{\zeta}$  & ${w(\zeta)} L^{\zeta}$\\
 $w(t,L)=$               & ${w_c(\zeta)}(1-e^{-t/\tau_c(\zeta)})$ & $L^{\zeta}\hat{w}(t/L^z)$ & $L^{\zeta}\hat{w}(t/L^z)$
\end{tabular}
\caption{Scaling functions of the observables considered both in the stationary and non-stationary cases, in the three scaling regimes.
\label{table}
}
\end{table}


In addition, we computed numerically the distribution of intervals $\pofl$ and the distribution of the number of zeros $P(n)$. For the distribution of intervals we found for regime $\RN{1}$ that $\pofl\approx e^{-\ell/\ell_c(\zeta)}G_{\RN{1}}(\zeta)$ with $G_{\RN{1}}(\zeta=-1/2)=1$ and $\ell_c(\zeta=-1/2)=1/\log 2$ as expected for an Ornstein--Uhlenbeck process. For regime~$\RN{2}$, the distribution of intervals was found to be $\pofl\approx \ell^{-2+\zeta} G_{\RN{2}}(\ell/L)$ with a power-law behavior for intermediate intervals and $G_{\RN{2}}(\ell/L)$ controlling the behavior for large intervals $\ell\sim L$. This result is consistent with the prediction $\theta=1-\zeta$~\cite{majumdar_spatial_2001}, implying $\gamma=\theta+1=2-\zeta$.
For regime $\RN{3}$, we find that $\pofl\approx \ell^{-3+2\zeta} G_{\RN{3}}(\ell/L)$ where the power-law behavior is found to be described by a different first-passage exponent than regime $\RN{2}$ for intermediate values of $\ell$, and  $G_{\RN{3}}(\ell/L)$ describes the behavior for large intervals proportional to the size of the system. This result has not been predicted before, and is different from the prediction $\gamma=\theta+1$  (valid for infinite interfaces). A possible reason for this discrepancy is the relevance of the zero-area constraint, not considered in Ref.~\cite{bray_persistence_2013}, 
for superrough interfaces. This prompts the question of why the same constraint does not equally affect the $\gamma$ exponent in the rough interface for regime \RN{2}. At this respect we also note that even for $\zeta=1/2$ the zero-area constraint is responsible for the breakdown of the independent interval approximation (a modified version of the Sparre-Andersen theorem is needed  for a correct description of $\pofl$ as studied in~\cite{zamorategui16distribution}). 

Regarding the distribution $P(n)$ of the number of zeros, it is less straightforward to draw  more accurate predictions. However, for regime $\RN{2}$ it was possible to measure heuristically the behavior of the distribution as follows. For $\zeta$ around $1/2$ and $n<\langle n\rangle$, we obtained that $P(n)\sim n^{2(1-\zeta)/\zeta}$. An analytical argument is still missing to validate such an exponent.

We consider that the results presented in this paper are both relevant theoretically and experimentally. On one hand we find that some properties of finite systems can be recovered from the results from non-constrained infinite-size systems. Understanding theoretically the role of constraints, such as the zero-area constraint, on the properties of rough interfaces and their associated persistence properties is a challenging task, in particular, for super-rough interfaces. It is worth stressing that these effects are not typical finite-size effects in the sense that they do not vanish in the large-size limit, since fluctuations themselves can scale even more rapidly with size (for instance $w_s \sim L^{2\zeta}$).
On the other hand, the effects due to the periodic boundary conditions, along with discretization effects are relevant for the practical analysis of zeros in interface imaging experiments or numerical simulations, as discussed below.

We believe many of our results should apply beyond the linear model we use.  
Indeed, in many non-linear cases, the geometry is rather well described by 
Gaussian statistics. One such case is the depinning transition of elastic interfaces in presence of quenched disorder, where the interface width can be accurately described by a Gaussian interface with a given roughness exponent~\cite{Rosso_contactline_2004}, in spite of the drastic modification of transport properties with respect to the linear model. Another case is the Kardar-Parisi-Zhang equation where a well-defined dynamical scaling is experimentally observed~\cite{Takeuchi_2010,Takeuchi2011}. 
Moreover, periodic boundary conditions make sense in practical situations where the interfaces are the perimeter of a static or growing droplet. One example is the case of magnetic bubbles in ferromagnets, which can be nucleated and driven by magnetic fields~\cite{Ferre2013}. Another example are bubbles of the turbulent phase in nematic liquid crystals~\cite{Takeuchi_2010,Takeuchi2011}.
Interestingly, both the steady-state and non-stationary results are relevant for growing droplets, as the dynamical length can grow faster or slower than the perimeter, depending on the drive and on the universal dynamical exponent $z$.


Regarding non-stationary stochastic systems where aging properties are experimentally observed, we believe that non-stationary persistence properties similar to the ones studied in the present paper can be analyzed. Such situation typically occurs when the relaxation is controlled by a growing dynamical length (below which the interface is locally stationary), smaller than the interface size. The non-stationary persistence properties we find may be also applied to non-linear interfaces too, provided their non-stationary dynamics is described by a Family--Vicsek scaling, with a dynamical exponent $z$ not necessarily related to $\zeta$ in the same way as in our model.  Such is the case of the relaxation of an initially flat interface at the depinning transition~\cite{Ferrero2013}, where an approximate Gaussian statistics is good enough to describe geometrical features as in Ref.~\cite{Rosso_contactline_2004}.
Although the growing dynamical length controlling the relaxation does not follow a power-law but a slower growth with time, as in transient creep motion or relaxation to equilibrium~\cite{kolton_relaxtoeq_2005}, we expect that our results can be applied.

There are many other interesting open questions to address.
In this paper we have only addressed the periodic boundary condition case, which is very convenient for numerical simulations and can be also realized experimentally. What differences can we expect from different boundary conditions? And in particular, what can we expect from the analysis of a finite segment of a large interface from an experiment? Such analysis has been already done for the width of Gaussian signals~\cite{santachiara_width_window2007}. Thus, it would be interesting to perform a similar analysis regarding the intervals between zeros. 
Another interesting question is the effects of non-linearities in the zeros statistics of self-affine interfaces. It is clear that non-linearities can drastically affect the dynamics of interfaces and change their universality class. 
More interestingly, they can produce geometrical crossovers (with different scales described by different roughness exponents). Understanding the effects of such crossovers in the statistics of zeros is also relevant. 
Finally, it would be interesting 
to validate the results obtained in this paper using analytical approaches, such as the perturbative method for non-Markovian Gaussian signals~\cite{delorme_2016,sadhu_generalized_2017}.

\section{\label{sec:conclusions}Conclusions}

In summary,  
we report a numerical study of 
the distribution of zero crossings 
in one-dimensional elastic interfaces described 
by an overdamped Langevin dynamics with periodic boundary conditions 
and fractional elasticity.
By continuously increasing from $\zeta=-1/2$ 
(macroscopically flat interface described by independent Ornstein--Uhlenbeck processes)
to $\zeta=3/2$ (super-rough Mullins-Herring interface), three 
different regimes are identified.
The results drawn from our analysis of rough interfaces subject to particular boundary conditions or constraints, along with discretization effects, are relevant for the practical analysis of zeros in interface imaging experiments or numerical simulations of self-affine interface models. 
 
\section{Acknowledgements}
We thank Gregory Schehr and Kazumasa Takeuchi for fruitful discussions.
The authors acknowledge the ECOS project A12E05 which allowed this collaboration. ABK wishes to thank the European Union EMMCSS visiting scientist program and 
acknowledges partial support from Projects PIP11220090100051 and PIP11220120100250CO (CONICET). VL acknowledges the CNRS Coopinter project EDC25533  and the CNRS PICS project 260693 for partial funding,
and support by the ANR-15-CE40-0020-03 Grant LSD and by the European Research Council (ERC) Starting Grant No. 680275 MALIG.
ALZ wishes to thank the hospitality of the Condensed Matter Theory Group at Centro At\'omico Bariloche, S.C. de Bariloche, Argentina, and acknowledges 
funding from the Mexican National Council for Science and Technology (CONACyT).

\bibliographystyle{apsrev}
\bibliography{Bibliography}

\pagebreak
\section{Appendix}
\subsection{General considerations on the zeros 
of truncated Fourier series}

In our study of zeros of interfaces with periodic boundary conditions we have used
truncated Fourier series
\begin{equation}
\label{eq:truncatedfourier0}
 u_x=\frac{1}{\sqrt{L}}\sum_{2\pi/L}^{q_{\text{cut}}} u_q e^{iqx},
\end{equation}
where $q_{\text{cut}}$ is an ``ultraviolet'' cut-off. For a discrete interface with $L$ elements the smallest cut-off is $q_{\text{cut}}=\pi$. However, we shall consider the general case, $q_{\text{cut}}=2\pi N/L$, which can arise simply from a smoothening of the discretized interface, yielding smaller values of $q_{\text{cut}}$.
We take $u_{q=0}=0$ in order to describe displacements around the mean position of the interface. 
Here we will make general considerations, regardless of the statistical properties of the amplitudes $u_{q>0}$. One might want to determine the minimum and the maximum number of zeros  that can be generated by the truncated series in Eq.~\eqref{eq:truncatedfourier0}. Since the interface displacement is real, $u_{q}=u_{2\pi -q}^*$,  
we can thus write
\begin{equation}
 u_t=\sum_{j=1}^{N} 
 [a_j \cos(j t) + b_j \sin(j t)]
\label{eq:trigonometricpolynomial}
\end{equation}
where $a_j$ and $b_j$ are real, and we have defined $N = L q_{\text{cut}}/2\pi$ and $t=2\pi x/L$. 
This trigonometric polynomial has exactly $2 N$ complex roots~\cite{Boyd2006}. It can be also proved that since $a_j$ and $b_j$ are real, the number of real roots is always even, if we count roots according to their multiplicity.
The maximum number of real roots is $2N$, which can be generated by setting $a_j=b_j=0$ except for $j=N$.  
A smaller number of real roots is possible, but since each of them have multiplicity two, it should be an even number too.
It is also clear that the fundamental mode alone, $j=1$ or $q=2\pi/L$, will generate the minimum of two real roots if the zero-area constraint, $a_0=0$, is imposed. Therefore, the maximum number of zeros $2N$ is set by the shortest wavelength $q_{\text{cut}}=2\pi N/L$ and the minimum should be $2$.

The above observations have two important consequences for the statistics of zeros: the zero-area constraint enforces $P(n)$ to have a hard cut-off at $n=2$, and $\pofl$ to have a soft cut-off at $\ell \sim L/2$. For $\zeta<0$ (regime $\RN{1}$), a typical configuration has $n\approx L/2$ zeros, or equivalently intervals with average length $\langle\ell\rangle\approx 2$. As $\zeta$ becomes positive, the amplitude of the modes with shorter wavelengths begin to be relevant and large intervals of the order of the system size are more likely to appear. This effect becomes more and more important as $\zeta$ increases, particularly in the super-rough regime \RN{3} for $\zeta>1$, where short-wavelength modes dominate the shape of typical configurations.

Moreover, the effect of smoothening by truncating the Fourier series imposes a hard cut-off in $P(n)$ at $n=2N$
having interesting consequences for the statistics of intervals between zeros, as discussed in the following Section.

In addition, if we define $y\equiv e^{it}$, with $t \equiv 2\pi x/L$, we can write an ``associated polynomial'' for $u_t \equiv u_{x = t L/2\pi}$ as
\begin{equation}
h(y)=\sum_{k=0}^{2N} h_k y^k = y^{N} u_{x(y)},
\end{equation}
where 
\begin{eqnarray}
h_k &=& a_{N-k}+i b_{N-k},\;\;\; k=0,1,...,N-1\nonumber \\
h_k &=& 2 a_0,\;\;\; k=N\nonumber  \\
h_k &=& a_{k-N}+i b_{k-N},\;\;\;j=N+1, N+2,..., 2N.
\quad
\end{eqnarray}
Therefore, the zeros of $u_t$ are those of $h(y)$. 
In our case, we should take $a_0=0$ to ensure the zero-area constraint.
This correspondence allows us to write the Fourier companion matrix
\begin{eqnarray}
B_{jk}&=&\delta_{j,k-1},\;\;\;j=1,2,...,(2N-1)\\
B_{jk}&=&-1 \frac{h_{k-1}}{a_N-i b_N}, \;\;\; j=2N
\end{eqnarray}
Interestingly, the roots of $u_{t}$ can thus be computed from the eigenvalues $\lambda_k$ of $B$ as~\cite{Boyd2006}:
\begin{equation}
t_{k,m}={\arg}(\lambda_k)+2\pi m - i \log |\lambda_k|,\;\;\; k=1,2,...,2N;\;\;\; \end{equation}
with $m$ an integer. If we restrict the zeros to $t \in [0,1]$ or $x \in \{0,1,...,L-1\}$ we can take $m=0$. 
Therefore, the real roots of the trigonometric polynomial are 
the complex eigenvalues of $B$ or the zeros of $h(y)$ in the complex-plane unit circle.
Thus, we recover the fact that $2N$ roots exist. Consequently, the connection with the associated polynomial provides us with a recipe to build a trigonometric polynomial with any desired even number of zeros, $n=2,4,6,...,2N$.

\subsection{Low-frequency modes analysis}
\label{app:low-modes}
In this section, we discuss the role that low-frequency modes have in the first-passage properties of the interfaces and smoothening effects. 
To do so we consider stationary interfaces described by a low number of Fourier modes, introducing a cut-off wavevector $q_{\text{cut}}$.  Thus, the interface described by the first $q_{\text{cut}}$ modes is defined as
\begin{equation}
 u_x=\frac{1}{\sqrt{L}}\sum_{2\pi/L}^{q_{\text{cut}}} u_q e^{iqx}
\label{eq:truncatedfourier}
\end{equation}
with $u_{q=0}=0$ (in order to describe displacements around the mean position of the interface) and $u_{q>0}$ defined with zero mean and second moment $\sigma^2_{q}=\langle u_q^2 \rangle=T/q^{1+2\zeta}$, 
in order to describe statistical self-affine interfaces with roughness $\zeta$.

Although our analysis can be implemented on interfaces with roughness exponent $\zeta\in(-1/2,3/2)$, in this section we focus on the particular case $\zeta=3/2$.

An interface described by two modes with the lowest frequencies, has the form
\begin{equation}
\label{eqtwomodes}
 u_x(t\to\infty)=u_x=\sin(q_1x)+\frac{u_{q_2}}{u_{q_1}} \sin(2q_1x+\phi_2).
\end{equation}
We investigate here the ratio of the random amplitudes $A=u_{q_2}/u_{q_1}$ and random phase $\phi_2\in(0,2\pi)$ that maximizes the number of intervals. Without loss of generality, we fix the phase of the first mode to $\phi_1=0$. As a matter of fact, the maximum number of intervals we can have for an interface described by Eq.~\eqref{eqtwomodes} is four, when the second mode dominates. The best scenario to generate a new zero $x_1$ close to $x_0=0$ is to fix the phase of the second mode to $\phi_2=\pi$. By fixing the phases, we can focus only on the ratio $A$ to understand how a new zero is created.

The ratio $A$ of two Gaussian variables with zero mean, is a random variable with a Cauchy distribution $f(A)=\frac{1}{\pi}\frac{r(\zeta)}{A^2+r(\zeta)^2}$, depending on a single parameter $r(\zeta)$ defined as the ratio of the mean squared values of the two modes: $r(\zeta)=\sigma_{q_2}/\sigma_{q_1}=2^{-(1+2\zeta)/2}$. The condition to have a new zero $x_1$ is that the slope at $x_0$ becomes negative, \emph{i.e.}~that $\partial_x u_x|_{x_0}<0$, which is the case for $A>1/2$. The same argument can be done to find a zero around $x=L/2$, for the value at which  $\partial_x u_x|_{x=L/2}>0$, which is the case for $A<-1/2$.

Let us consider, for instance, an interface with $\zeta=3/2$ (for which $r(\zeta)$ reaches its minimum value) described by only two modes with relative phase $\pi$. In this case, the probability of observing four zeros is $F(|A|>1/2)=1-F(|A|\leq1/2)\approx 0.3$. In this expression, $F(A\leq b)=\int_{-\infty}^{b} f(A')dA'$ is the cumulative distribution function, equal to $F(A)=\frac{1}{\pi}\arctan(A/r(\zeta))+\frac{1}{2}$.
 
 \begin{figure}[ht!]
 \centering
 \includegraphics[scale=1.1]{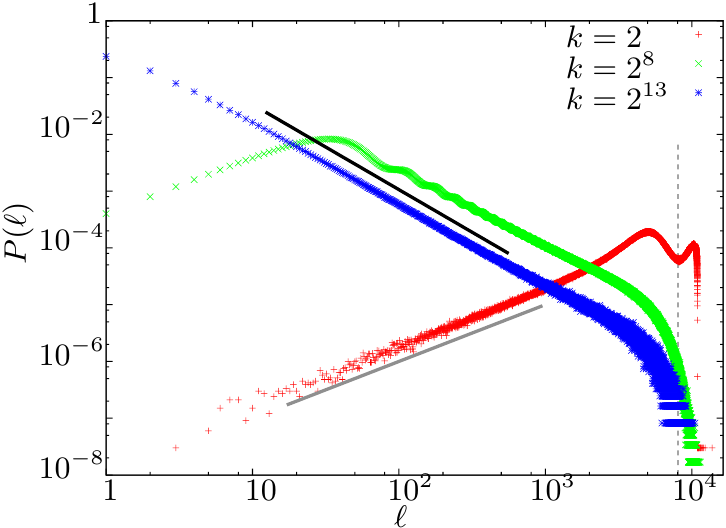}
 \caption{Distributions of intervals $P(\ell)$ for $\zeta=1/2$ and $L=16384$ obtained for different values of $q_{\text{cut}}=2\pi k/L$ with $ k=2,256,8192$ (see Eq.~\eqref{eq:truncatedfourier}). The solid lines correspond, respectively, to the power-law regime $\ell^{-3/2}$ (black line) and the linear behavior (gray line) discussed in this section. The vertical line pinpoints the value $L/2$.  \label{twomodesforzetaonehalf}}
\end{figure} 

 \begin{figure}[ht!]
 \centering
 \includegraphics[scale=1.1]{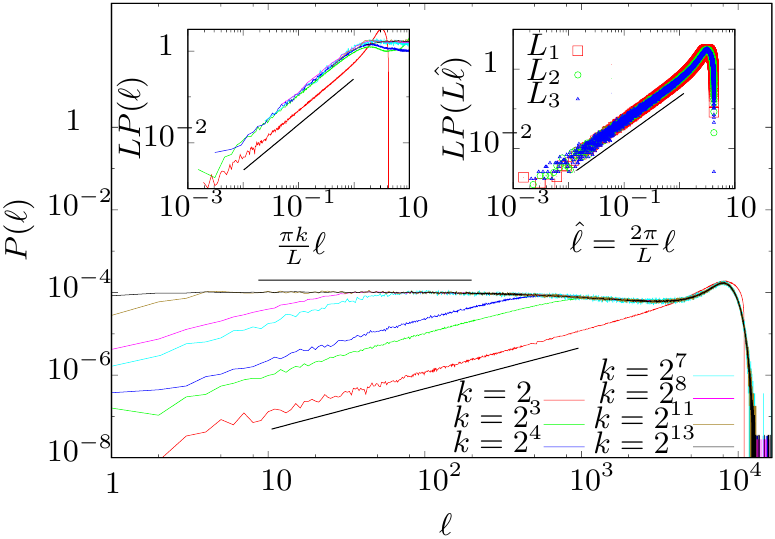}
 \caption{Distributions of intervals $P(\ell)$ for $\zeta=3/2$ and $L=16384$ obtained for different values of $q_{\text{cut}}=2\pi k/L$ with $ k=2,8,16,128,256,2048,8192$ (see Eq.~\eqref{eq:truncatedfourier}). The diagonal line shows the limiting behavior $\pofl\sim \ell$ for small intervals of size $\ell<L/2 k$. For $\ell>L/2 k$, $\pofl$ remains unchanged for $ k>2$. As  shown in the inset on the left, this dependence on $q_{\text{cut}}$ validates the behavior found analytically for the regime of $\ell<L/2 k$ where the scaling $P(\ell)=\frac{1}{L}\hat{P}(\ell q_{\text{cut}})$ collapses the linear behavior for $ k\geq 8$. Such a linear behavior is exactly predicted for $ k=2$ as obtained from Eq.~\eqref{distributionelltwomodes} and depicted by the solid line. The scaling behavior of $\pofl$ for different $q_{\text{cut}}$ indicates the transition at $\ell\sim 1/q_{\text{cut}}$ from the linear behavior for small $\ell$ to the regime given by the plateau whose height is of order $\sim 1/L$.  The inset on the right validates the scaling of the interval distribution $\pofl$ described by the two first low-frequency modes given by the expression $LP(\ell)\sim \frac{2 \pi \ell}{ L}$, for different system sizes $L$: $L_1=16384$, $L_2=32768$ and $L_3=65536$ (See Eq.~\eqref{distributionelltwomodes} and the discussion that follows).  \label{nqcut}}
\end{figure}

 Then, the new zero $x_1$ around $x_0$, is the solution of $\sin(2\pi x_1/L)=A\sin(4\pi x_1/L)$, where $A>1/2$, thus
\begin{equation}
\label{xofA}
 x_1=\frac{L}{2\pi}\arctan(\sqrt{-1+4A^2}).
\end{equation}
We note that the value of $x_1$ gets close very quickly to $L/4$ as soon as $A>2$. In conclusion, this leaves us with two intervals of length $\ell_1=x_1$ and two intervals of length $\ell_2=\tfrac{L}{2}-\ell_1=\tfrac{L}{2}(1-\tfrac{\arctan(\sqrt{-1+4A^2}}{\pi})$.  

From Eq.~\eqref{xofA}, we can also express $A$ as a function of the position of the zeros. Then, from the Cauchy probability $f(A(x))$ we obtain the distribution of $x$ as
\begin{equation}
\label{distributionelltwomodes}
 P(x)=f(A(x))\frac{dA}{dx}=\frac{4r\sin(2\pi x/L)}{L(1+2r(\zeta)^2+2r(\zeta)^2\cos(4\pi x/L))},
\end{equation}
where $r(\zeta)=2^{-(1+2\zeta)/2}$. Since the first zero of the interface was fixed to $x_0=0$, the first interval has a length $\ell=x$. Hence, its distribution is 
\begin{equation}
\label{distributionsmallell}
 P(\ell)\approx \frac{8 \pi r \ell}{ L^2(1+4r(\zeta)^2)}+O(\ell^3),
\end{equation}
which is obtained from Eq.~\eqref{distributionelltwomodes} for small $\ell<L/4$. In Figs.~\ref{twomodesforzetaonehalf} and \ref{nqcut}  we computed numerically $\pofl$ for a system of size $L=16384$ and $\zeta=1/2$ and $\zeta=3/2$, respectively, for different values of $q_{\text{cut}}=2\pi k/L$, with $ k$ the number of non-trivial modes taken into account. In particular, the curves corresponding to $q_{\text{cut}}=4\pi/L$ (thus $ k=2$) in Figs.~\ref{twomodesforzetaonehalf} and~\ref{nqcut}, validate the linear behavior derived in Eq.~\eqref{distributionsmallell}.

The linear behavior in Eq.~\eqref{distributionsmallell} is observed for any roughness exponent, with only a prefactor that depends on $\zeta$ and the cut-off $q_{\text{cut}}$. Here we derived the exact expression of this prefactor for the case $q_{\text{cut}}=4\pi/L$, corresponding to having two Fourier modes. 

Further, the linear regime~\eqref{distributionsmallell} is ubiquitous for any $q_{\text{cut}}$ and any roughness exponent $\zeta$ provided that the length of the intervals satisfies $\ell<1/q_{\text{cut}}$. For intervals where $\ell\gtrsim 1/q_{\text{cut}}$ the distribution of intervals observed is close to the complete distribution $P(\ell)$ when all the modes are considered. It is still an open question to understand how the power-law regime of $P(\ell)$ develops, in general, and how the plateau appears for $\zeta=3/2$, in particular.

The exact determination of the length of the intervals for an interface reduced to two Fourier modes gives some insight in the way the zeros of the interface are generated. In general, the presence of a large interval is accompanied by a small interval. This is the case when the relative phases are close to $\pi$ and four zeros are produced. In this case, either we observe either four intervals with lengths close to $L/4$, or two large intervals with length $\ell\lesssim L/2$ with two very small intervals in between. This fact is captured by the correlation of intervals, as shown in Appendix \ref{app:correlations}.  Although the analysis of such correlations goes beyond the goal of this article, we provide the results of their numerical evaluation for completeness.

\subsection{Correlation of intervals.}
\label{app:correlations}
The interfaces we study are strongly correlated due to the periodic boundary conditions. The zero-area constraint is equivalent to having the sum of the individual increments along the interface equal to zero. Thus, both the intervals and the increments are correlated at all distances. In~\cite{zamorategui16distribution} we showed that for the EW model ($\zeta=1/2$), not only consecutive increments but at any distance are correlated.

\begin{figure}[ht!]
 \centering
     \subfigure[][]{\includegraphics[scale=0.9]{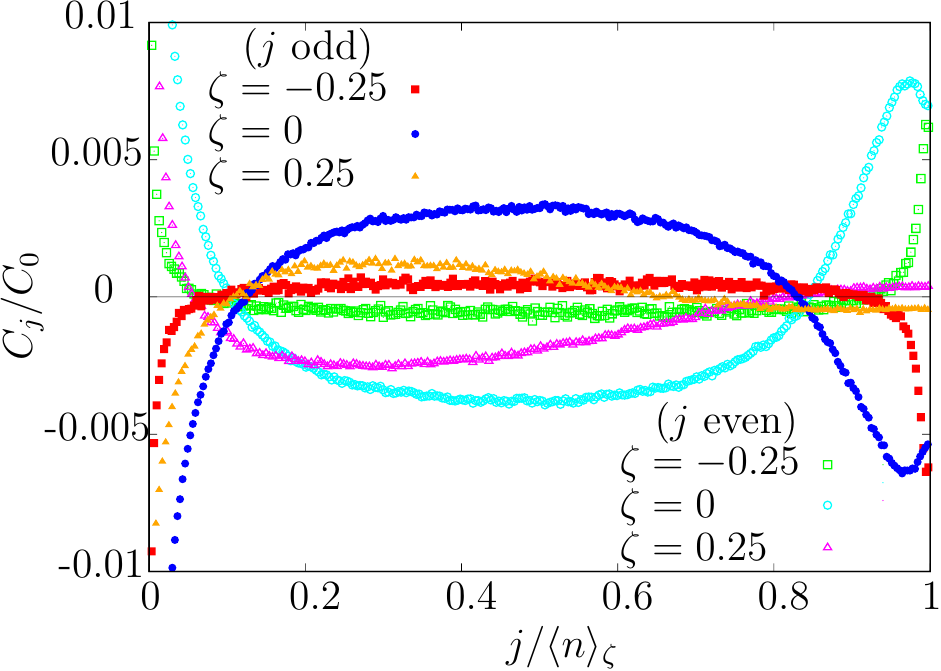}}
  \subfigure[][]{\includegraphics[scale=0.9]{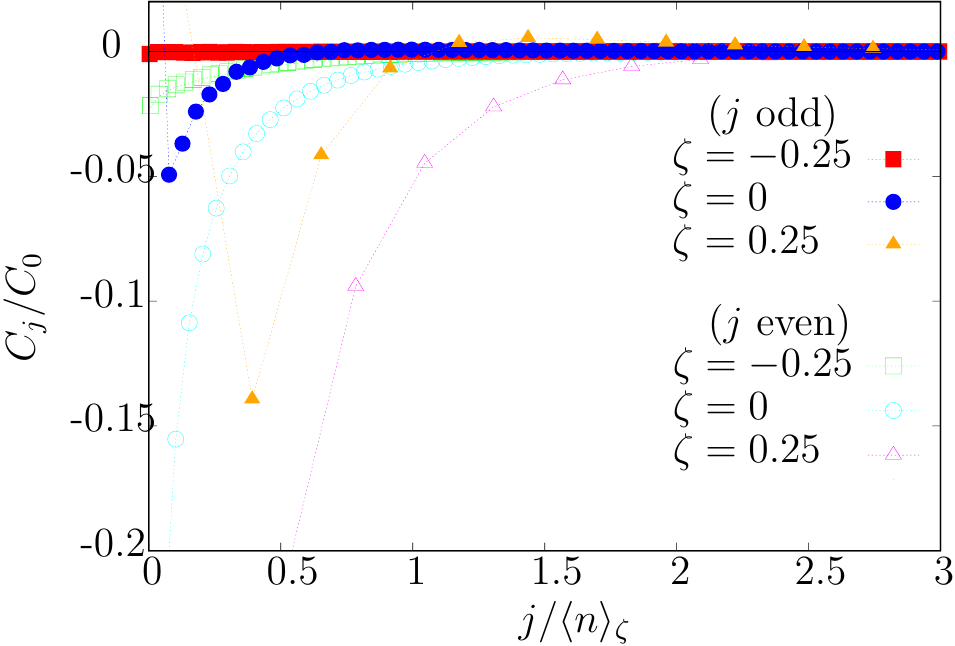}}
 \caption{ Correlations $C_j/C_0$ of intervals for different $\zeta$ with the correlation function defined as $C_j=\langle \ell_i\ell_{i+j}\rangle_i$ as a function of the $j$-th interval normalized by the average number of intervals $\langle n\rangle_{\zeta}$. We compare $C_j$ for $j$ even and odd. For $\zeta<1$ we observe that $C_j$ for even $j$ (for odd $r$)  goes from correlated (anti-correlated) to anti-correlated (correlated) intervals. Then such process is reversed and gets peaked around $\langle n \rangle$ which corresponds to the mean value of the number of zeros (thus of the number of intervals). Such mean value coincides with the maximum value of the distribution $P(n)$ in Fig.~\ref{pofnzerosall}. The intervals $\ell$ are completely decorrelated for $\zeta=-1/2$ since this process corresponds to a random coin toss process that takes positive or negative values, independently at each site. Hence, for such a process an interval of length appears with probability $P(\ell)=(1/2)^{\ell}$, as discussed in the text.  The value of $|C_j|$ decreases as $\zeta$ becomes larger.   \label{intcorr}}
\end{figure}

Regarding the correlations of the intervals, we define the correlation function $C_j=\langle \ell_i\ell_{i+j}\rangle-\langle \ell \rangle^2$ where $\langle\cdot \rangle$ refers to the average taken over all intervals. In Fig.~\ref{intcorr} we show such correlation for $\zeta<1/2$ and $\zeta>1/2$, respectively. Above $\zeta=1/2$, the correlation function both for even and odd neighbors is, in general, anti-correlated, except for small regions. This might be explained by the analysis done in Appendix \ref{app:low-modes} where we get the exact length of the intervals generated by the first two non-trivial modes: large intervals are followed by small intervals. If we take into account all the modes, then the probability of having small intervals increases. 

\subsection{Exact results for the interface width}
\label{app:widthanalytic}
We derive here exact predictions for stationary and non-stationary interfaces with mean squared width defined as
\begin{eqnarray}
w^2(t) &=& L^{-1} \sum_x \langle u_x^2 \rangle .
\end{eqnarray}


Our starting point is the discrete space $x=0,...,L-1$, 
continuous time spatially fractional equation:
\begin{eqnarray}
  \partial_t u_x = -(-\partial^2_x)^{z/2} u_x + \eta_x(t), \\
  \langle \eta_x(t) \eta_{x'}(t') \rangle = 2 T \delta_{x,x'} \delta(t-t')
\end{eqnarray}
where $z \equiv 1+2\zeta$. We use periodic boundary conditions 
so $u(x)$ is a spatially periodic function with period $L$.
Let us fix the forward and backward transform conventions:
\begin{eqnarray}
  u_q &=& \sum_x e^{-i q x} u_x(t)/\sqrt{L} \\
  u_x &=& \sum_q e^{i q x} u_q(t)/\sqrt{L}
\end{eqnarray}
with $q=2\pi k/L$, and $k=0,1,...,L-1$. 
Then, in Fourier space, due to the 
translational symmetry of the dynamics, 
all modes are uncoupled
\begin{eqnarray}
  \partial_t u_q = -|q|^z u_q + \eta_q(t), \\
  \langle \eta_q(t) \eta_{q'}(t') \rangle = 2 T \delta_{q,-q'} \delta(t-t')
\end{eqnarray}

If we start with a flat initial condition at the origin, $u_q(t=0)=0$, the 
non-stationary solution is, for $q\neq 0$
\begin{eqnarray}
u_q(t) = \int_0^t dt'\;\exp[-|q|^z (t-t')] \eta(t'). 
\end{eqnarray}
Since we will be interested in displacements with respect to the center of mass of the interface we will consider
$u_{q=0}(t)=0$ $\forall\; t$. 


The structure factor, defined as 
$S_q(t) \equiv \langle u_q(t) u_{-q}(t) \rangle = \langle |u_q(t)|^2 \rangle$, 
is
\begin{eqnarray}
S_q(t) =
\int_0^t \int_0^t dt_1 dt_2 \;
e^{-|q|^z (2t-t_1-t_2)}
\langle  \eta(t_1)\eta(t_2) \rangle.
\end{eqnarray}
Then,  for $q\neq 0$ 
\begin{eqnarray}
S_q(t) = T|q|^{-z} \left[ 1-\exp(-2|q|^z t) \right]
\end{eqnarray}

The mean squared width growth of the interface or roughness is then
\begin{eqnarray}
w^2(t) &=& L^{-1} \sum_x \langle u_x^2 \rangle \nonumber \\ 
&=&  L^{-2} \sum_x \sum_q \sum_{q'} \langle u_q u_{q'} \rangle \exp[i (q+q') x] 
\nonumber \\ 
&=&  L^{-1} \sum_q \langle |u_q(t)|^2 \rangle 
= L^{-1} \sum_q S_q(t) \nonumber  \\ 
&\approx& 2TL^{-1} \int_{2\pi/L}^{\pi} \frac{dq}{2\pi/L} \; |q|^{-z} \left[ 1-\exp(-2|q|^z t) \right] \nonumber \\ 
&=&
\frac{T}{\pi} \int_{2\pi/L}^{\pi} {dq} \; |q|^{-z} \left[ 1-\exp(-2|q|^z t) \right]
\end{eqnarray}


We now analytically evaluate the last expression in different limits.

\paragraph{Large time limit ($t\to \infty$).}
If $z\neq 1$ ($\zeta \neq 0$) we have
\begin{eqnarray}
w^2(t \to \infty) &\approx& \frac{T}{\pi} \int_{2\pi/L}^{\pi} dq\; |q|^{-z} 
\nonumber \\
&=& \frac{T}{\pi (1-z)} 
\left[\pi^{1-z}-\left(\frac{L}{2\pi}\right)^{z-1}\right]
\label{eq:largetimes}
\end{eqnarray}
We thus observe the following
\begin{itemize}
  \item 
In the large size limit, if $\zeta<0$ ($z<1$), $w^2(t \to \infty)$ is dominated by the ultraviolet cut-off 
\begin{eqnarray}
w^2(t \to \infty)\approx \frac{T}{1-z} \pi^{-z}.
\label{eq:largetimeslargesizenegative}
\end{eqnarray}
If $z=0$, $w^2(t \to \infty)=T$ as expected from the energy equipartition-theorem, since $z=0$ 
corresponds to $L$ uncoupled overdamped Langevin oscillators (with a Hooke spring constant equal to the unity). 
\item For $\zeta>0$ it is dominated by the infrared size dependent cut-off,
\begin{eqnarray}
w^2(t \to \infty)&\approx& 
\frac{2T}{2\pi (z-1)} \left(\frac{L}{2\pi}\right)^{z-1} 
\nonumber \\
&=&
\frac{2T}{(2\pi)^z (z-1)} L^{2\zeta} 
\label{eq:largetimeslargesizepositive}
\end{eqnarray}

\item In the marginal case $z=1$, $\zeta=0$, we have
\begin{eqnarray}
w^2(t \to \infty)\approx \frac{T}{\pi} \int_{2\pi/L}^{\pi} dq\; |q|^{-1} 
= \frac{T}{\pi} \log(L/2)
\end{eqnarray}
\end{itemize}

\paragraph{Small times}
\begin{eqnarray}
w^2(t) &\approx& 
\frac{T}{\pi} \int_{2\pi/L}^{\pi} {dq} \; |q|^{-z} \left[ 1-\exp(-2|q|^z t) \right] 
\nonumber \\
&\approx& 
\frac{T}{\pi} \int_{2\pi/L}^{\pi} {dq} 2t 
= 
{2Tt} (1-2L^{-1}) \approx 2Tt 
\label{eq:smalltimes}
\end{eqnarray}
and it holds for any $z$. Small 
times have thus always diffusive displacements, independent of $z$.

\begin{figure}[ht!]
 \centering
\includegraphics[scale=0.9]{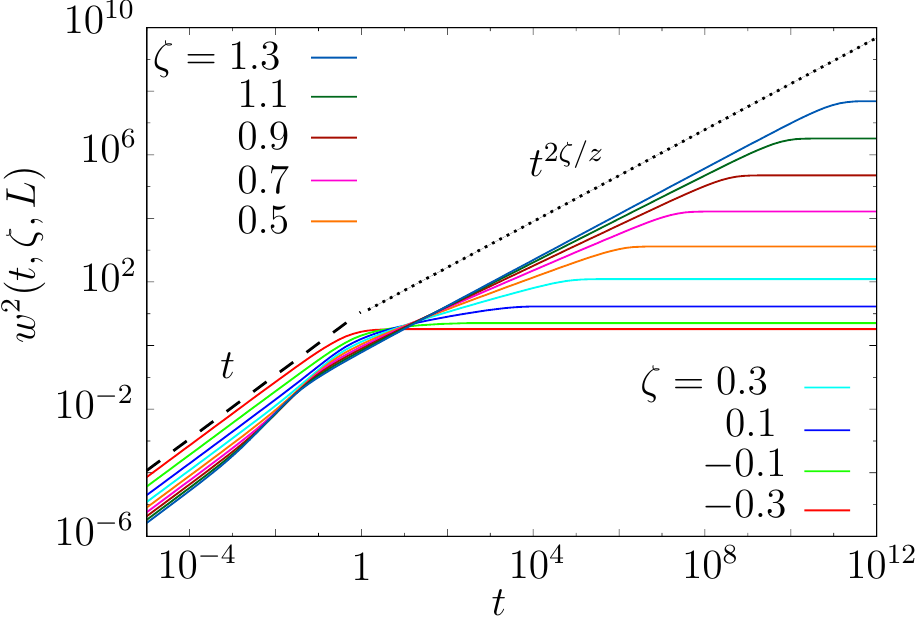}
\caption{Time evolution of the mean squared interface width $w^2(t,\zeta,L)$ for different values of $\zeta$ (or $z=1+2\zeta$). The initial condition is a flat interface with $w^2(t=0)=0$ and the zero area constraint is imposed at all times. The dashed line indicates the diffusive behavior $w^2(t)\sim t$ at short times for all $\zeta$. At larger times, the dotted line indicates the 
critical behavior $w^2(t)\sim t^{2\zeta/z}$
expected for $\zeta>0$ (here we show the case $\zeta=1.3$), before saturating to 
$w^2(t\to \infty)\sim L^{2\zeta}$
at even larger times $t\sim L^z$.
For $\zeta<0$ ($z<1$) the diffusive regime crossovers directly to an $L$-independent saturation but at $\zeta=0$ ($z=1$), $w^2(t\to \infty)\sim \log L$.}
\label{fig:wvstanal}
\end{figure}

\paragraph{Finite times}
If $z=0$ we can be 
readily solve for all times $t \geq 0$
\begin{eqnarray}
w^2(t) &\approx& 
\frac{T}{\pi} \int_{2\pi/L}^{\pi} {dq} \; |q|^{-z} \left[ 1-\exp(-2|q|^z t) \right]
\nonumber \\
&=& \frac{T}{\pi} \left[ 1-\exp(-2 t) \right] \pi(1-2 L^{-1}) 
\nonumber \\
&\approx& {T} \left[ 1-\exp(-2 t) \right] 
\end{eqnarray}
where in the last term we took the large size limit, $L \to \infty$. It is 
worth noting however that 
relaxation is not critical and it has a finite size-independent characteristic time.

%
In the more general $z\neq 0$, $z\neq 1$ we can also analytically solve and obtain


\begin{eqnarray}
w^2(t) 
&\approx& 
t^{2\zeta/z} [F(\pi t^{1/z},z)-F(2\pi L^{-1} t^{1/z},z)],
\label{eq:finitetimes}
\end{eqnarray}
where 
\begin{eqnarray}
F(y,z)\equiv 
\frac{(e^{-2 y^z}-1) y^{1-z} - 2^{\frac{z-1}{z}} \Gamma[z^{-1}, 2 y^z]}{z-1} 
\end{eqnarray}
with $\Gamma[a,x]$ is the ``upper'' incomplete gamma function.
With the above expression we find that the mean squared width $w^2(t)$ displays two marked different behaviors.

\begin{itemize}
\item
If $z<1$, 
$w^2(t)$ initially increases linearly with time as predicted in Eq.~\eqref{eq:smalltimes} and in an $L$-independent finite time crossovers, exponentially fast, towards a $z$-dependent stationary value described by Eq.~\eqref{eq:largetimeslargesizenegative} in the large $L$ limit. This regime is thus size independent in such a limit.
\item
For $z>1$ on the other hand, $w^2(t)$ also initially increases linearly with time 
as predicted in Eq.~\eqref{eq:smalltimes} 
and in a finite $L$-independent time crossovers, exponentially fast, to a different power law growth, $w^2(t) \sim t^{2\zeta/z}$. This growth regime in turn crossovers exponentially fast, after a time that scales as $L^z$,  to the $L$ and $z$ dependent stationary value of Eq.~\eqref{eq:largetimeslargesizepositive}.  
This regime is thus controlled by infrared divergences.
\end{itemize}
In Figure \ref{fig:wvstanal} we show this behavior for different values of $z$, by plotting 
Eq.~\eqref{eq:finitetimes}.

\end{document}